\documentclass[english,12pt,aps,prd,a4paper,preprintnumbers,floatfix,nofootinbib,superscriptaddress, notitlepage]{revtex4} 
\usepackage{graphicx}
\usepackage[utf8]{inputenc} 
\usepackage{amsmath}
\usepackage{amssymb}
\usepackage{float}
\usepackage{comment}
\usepackage{placeins} %%for FloatBarrier
\usepackage{slashed}
\usepackage[normalem]{ulem}
\usepackage{braket}
\usepackage{makecell}
\usepackage{multirow}
\usepackage{caption}
\usepackage{tabularx}
\usepackage{adjustbox}
\usepackage[usenames,dvipsnames]{color}
\usepackage[colorinlistoftodos]{todonotes}
\usepackage[colorlinks=true,citecolor=darkred,urlcolor=darkred, pdfborder={0 0 0}]{hyperref}
\definecolor{darkred}{rgb}{0.6,0,0}
\newcommand {\ignore}[1]{}

\newcommand{\bea}{\begin{eqnarray}}
\newcommand{\eea}{\end{eqnarray}}
\usepackage{amstext}
\def\gsim{\raise0.3ex\hbox{$\;>$\kern-0.75em\raise-1.1ex\hbox{$\sim\;$}}}
\def\lsim{\raise0.3ex\hbox{$\;<$\kern-0.75em\raise-1.1ex\hbox{$\sim\;$}}}

\usepackage{soul}

\definecolor{mightnightblue}{RGB}{25,25,112}

\definecolor{brown}{rgb}{0.59, 0.29, 0.0}

\DeclareUnicodeCharacter{2032}{'}
\DeclareUnicodeCharacter{03BC}{\ensuremath{\mu}}

\def\21{$\mathrm{SU(2)_L \otimes U(1)_Y}$}

\usepackage{mathrsfs}
\usepackage{amsmath}
\usepackage{orcidlink}

\newcommand{\AddrIISERB}{Department of Physics, Indian Institute of Science Education and Research - Bhopal, \\ 
Bhopal Bypass Road, Bhauri, Bhopal 462066, India}
\begin{document}

%%%%%%%%%%%%%%%%%%%%%%%%%%%%%%%%%%%%%%%%%%%%%%%%%%%%%%%%%%%%%%%%%%%%%%%%
\title{\textcolor{BrickRed}{Dark hypercharge Symmetry}}
%%%%%%%%%%%%%%%%%%%%%%%%%%%%%%%%%%%%%%%%%%%%%%%%%%%%%%%%%%%%%%%%%%%%%%%%%
\author{Hemant Prajapati~\orcidlink{0000-0001-5104-9427}}\email{hemant19@iiserb.ac.in}
\affiliation{\AddrIISERB}
\author{Rahul Srivastava~\orcidlink{0000-0001-7023-5727}}\email{rahul@iiserb.ac.in}
\affiliation{\AddrIISERB}
%\date{\today}

%%%%%%%%%%%%%%%%%%%%%%%%%%%%%%%%%%%%%%%%%%%%%%%%%%%%%%%%%%%%%%%%%%%%%%%%%%%%%%%%%%%%%
\begin{abstract}
%%%%%%%%%%%%%%%%%%%%%%%%%%%%%%%%%%%%%%%%%%%%%%%%%%%%%%%%%%%%%%%%%%%%%%%%%%%%%%%%%%%%%
\vspace{0.5cm}

We introduce a new class of $U(1)_X$ symmetries where all Standard Model fermions are ``chiral," i.e. the left- and right-handed components have different charges under the $U(1)_X$ symmetry. Gauge anomaly cancellation is achieved by introducing three Standard Model gauge singlet dark fermions ($f^i$; $i=1,2,3$) charged under this symmetry. We systematically present chiral solutions for cases in which (a) one, (b) two, or (c) all three generations of Standard Model fermions are charged under the $U(1)_X$ symmetry. The $U(1)_X$ charges of these dark fermions are uniquely determined by anomaly cancellation conditions. These new fermions belong to the dark sector, with the lightest of them being a good dark matter candidate. Additionally, the $Z'$ gauge boson mediates interactions between the dark and visible sectors, and we call this $U(1)_X$ symmetry as the ``dark hyperCharge" symmetry. Using a benchmark model, we explore phenomenological implications in the heavy $Z'$ case ($M_{Z'} > M_Z$), analyzing collider constraints and examining the lightest dark fermion's viability as dark matter. Our analysis shows that it satisfies all current dark matter constraints over a wide range of dark matter mass.
%%%%%%%%%%%%%%%%%%
\end{abstract}
%%%%%%%%%%%%%%%%%%
\maketitle
%%%%%%%%%%%%%%%%%%

\section{Introduction}
\label{sec:intro}

The Standard Model (SM) is a highly successful theory, with most of its predictions verified by experiments. Despite its ability to explain most observable phenomena, the SM cannot be considered the final theory of nature.
An ever-increasing number of observations, such as the discovery of neutrino oscillations \cite{Super-Kamiokande:1998kpq,SNO:2002tuh} and the existence of dark matter (DM) at cosmic scales \cite{Zwicky:1933gu,Rubin:1970zza,Rubin:1980zd,Planck:2018vyg}, highlight issues that cannot be resolved within the framework of the SM. Thus, the SM needs to be extended by some beyond Standard Model (BSM) physics which can explain these shortcomings.
One of the most popular approaches is to extend the SM gauge symmetries by addition of new gauge symmetries. Among these extensions, the simplest and highly motivated one is an extra $U(1)_{X}$ gauge symmetry, a selected few works on such extensions can be found in Refs.~\cite{He:1990pn,Ma:1997nq,Appelquist:2002mw,Montero:2007cd,Lee:2010hf,Ma:2014qra,Ma:2015raa,Ma:2015mjd,Das:2016zue,Bonilla:2017lsq,Alonso:2017uky,Jana:2019mez, DeRomeri:2023ytt,Mandal:2023oyh,Ghosh:2024cxi}. 

The introduction of new gauge symmetries, particularly $U(1)_{X}$, requires careful consideration as it can lead to the emergence of gauge anomalies \cite{Adler:1969gk,Bardeen:1969md,Bell:1969ts,Delbourgo:1972xb,Alvarez-Gaume:1983ihn, Witten:1982fp}. Ensuring gauge anomaly cancellation is essential for maintaining unitarity and renormalizability in gauge theories. These gauge anomalies are related with the representation of fermions under gauge symmetry. If a fermion is vectorlike, i.e., both its left- and right-handed components share the same representation under the gauge symmetry, then it will not contribute to gauge anomalies in the theory. Conversely, if a fermion is chiral, i.e., its left- and right-handed components have different representations under the gauge symmetry, then it can introduce gauge anomalies. These anomalies must cancel out in a consistent gauge theory when considering contributions from multiple chiral fermions. Hence, anomaly cancellation imposes constraints on the possible charges of chiral fermions. 

The SM is also a chiral theory, with quarks and leptons being chiral under both $SU(2)_{L}$ and hypercharge $U(1)_{Y}$ gauge symmetries. As hypercharge is an Abelian gauge symmetry, this implies that the gauge anomalies in the SM need to be canceled carefully. The conditions for anomaly cancellation in the SM can be expressed in terms of the hypercharges of the SM fermions. These hypercharges add up in such a way that the anomalies cancel out. However, as we discussed at length in later sections, anomaly cancellation conditions alone are not sufficient to determine the hypercharge assignments of SM fermions uniquely. Additional constraints come from the requirement that weak isospin and hypercharge combine to yield the correct electric charge. Also, chiral SM fermions must also generate masses through Yukawa couplings with the Higgs boson. These two conditions, alongside gauge anomaly cancellation conditions, uniquely determine the hypercharges in the SM. 

The introduction of new $U(1)_{X}$ gauge symmetries places additional constraints on the possible charge assignments for both SM and BSM fermions. Moreover, if SM fermions are chiral under this $U(1)_{X}$ symmetry, then their mass and mixing generation mechanism may impose further constraints on the BSM charge assignments of these fermions. In the context of BSM physics, where new $U(1)_{X}$ symmetries are introduced, it is commonly assumed that all fermions are vector under the $U(1)_{X}$ symmetry. This assumption drastically simplifies the anomaly cancellation conditions and allows one to find viable anomaly-free models easily. Furthermore, vector charge assignment typically leaves most of the SM Yukawa coupling invariant under $U(1)_{X}$, thereby allowing the generation of SM fermion masses via the Higgs mechanism. \footnote{When $U(1)_{X}$ symmetry is generation dependent then additional scalars may be needed for generating masses and mixings of SM fermions.} 
Such vector solutions have been extensively studied.  Some notable examples are $B-L, B-3L_{i}, B_{i}-3L_{j}, L_{i}-L_{j} $, as well as their linear combinations, where $i,j= 1,2,3$ denote the generation index \cite{Ma:1997nq,He:1990pn,Appelquist:2002mw, Lee:2010hf,Ma:2014qra,Ma:2015raa,Ma:2015mjd,Bonilla:2017lsq,Alonso:2017uky, DeRomeri:2023ytt,Ghosh:2024cxi}. In addition in Sec.~\ref{SubSec:U(1)x_vector_solu} we have briefly discussed the possibility of new vector solutions   $B_{i}-B_{j}$; $i,j = 1,2,3$ are quark generation indices, which to the best of our knowledge is not yet discussed in the literature.

Apart from vector solutions, one can have another class of solutions, namely the chiral solutions, where the charges of left- and right-handed fermions differ from each other under the $U(1)_{X}$ symmetry. For chiral solutions, the $U(1)_{X}$ charges can be chosen in such a way that the overall anomalies cancel out. The hypercharge in the SM serves as an example of such a chiral solution. The chiral charge assignment offers a wide range of possible solutions, but only a few of them have been explored in the literature  \cite{Appelquist:2002mw,Montero:2007cd,Ma:2014qra,Das:2016zue,Jana:2019mez}.

In this study, we systematically investigate potential chiral solutions and introduce entirely new classes of solutions for (a) one, (b) two, and (c) three generations of SM fermions that are charged under the $U(1)_X$ symmetry.
The anomaly cancellation conditions for such solutions require the addition of new fermions, which are SM gauge singlets. The $U(1)_X$ charges of these BSM fermions are also fixed by the anomaly cancellation conditions. 
We show that these BSM fermions belong to the dark sector and the lightest of them is a good DM candidate. 
Thus,  the $Z'$ gauge boson associated with these chiral $U(1)_X$ symmetries provides the connection between the dark sector and the visible sector. 
Hence, we refer to these chiral $U(1)_X$ symmetries as dark hypercharge (DHC) symmetries.

The paper is organized as follows:
in Sec. \ref{Sec:Hypercharge_Anonaly_Uniqueness}, we discuss gauge anomaly cancellation conditions within the SM and examine the uniqueness of hypercharge assignments. Section \ref{sec:u1x-anomaly} then explores the conditions required for gauge anomaly cancellation in $U(1)_{X}$ extensions of the SM. This section also provides an in-depth analysis of various vector and chiral solutions designed to achieve gauge anomaly cancellation. In Sec. \ref{Sec:U(1)x_DarkHypercharge_solu}, we introduce a new class of chiral solutions, termed DHC symmetries. Section \ref{Sec:DHC_Gauge_Boson} then presents a generalized framework for the generation of the $Z'$ boson mass, achieved by incorporating SM singlet scalars $\chi_{i}$, and includes a thorough analysis of the $\rho$ parameter. In Sec. \ref{Sec:Zpcollider}, we examine $Z'$ boson production and decay channels for a benchmark DHC symmetry, evaluating the constraints imposed on this symmetry by LHC $Z'$ searches. In Sec. \ref{Sec:DM} we show that the lightest dark sector fermion can indeed be a good dark matter candidate. We summarize and discuss our main conclusions in Sec. \ref{Sec:Conclusion}. 
We also present some key details of this work in the Appendixes. Appendix \ref{APn3} details the partial decay width of the $Z'$ boson, relevant to discussions in this paper, while Appendix \ref{APn1} provides Feynman diagrams contributing to DM relic density and direct detection.
%
%%%%%%%%%%%%%%%%%%%%%%%%%%%%%%%%%%%%%%%%%%%%%%%%%%%%%%%%%%%%%%

\section{Anomaly Cancellation in SM: Uniqueness of Hypercharge}
\label{Sec:Hypercharge_Anonaly_Uniqueness}
%%%%%%%%%%%%%%%%%%%%%%%%%%%%%%%%%%%%%%%%%%%%%%%%%%%%%%%%%%%%%%

Let us begin with a discussion of how anomalies cancel in SM. This will help us set up the problem and notation for the rest of the paper.
Before discussing the anomalies of SM, let us first introduce a crucial nomenclature for fermions: 
\begin{itemize}
    \item \textbf{Vector fermions:}  We call a fermion ($\psi$) a \textbf{vector} fermion under a given gauge symmetry if its left- $(\psi_{\mathtt{L}})$ and right-handed $(\psi_{\mathtt{R}})$ components are in the same representation under the symmetry. In such cases, the invariant Dirac mass term for the fermions, i.e.  $m\, \bar{\psi}_{\mathtt{L}} \psi_{\mathtt{R}} + h.c.$ is allowed by the gauge symmetry. If there are multiple gauge symmetries, then such a mass term is allowed only if a fermion is a vector under all the symmetries.  Vector fermions do not induce anomalies related with the gauge symmetry under which they are vector.  

\item \textbf{Chiral Fermions:} A fermion ($\psi$) is called a \textbf{chiral} fermion under a given symmetry if its left- $(\psi_{\mathtt{L}})$ and right-handed $(\psi_{\mathtt{R}})$ components are in different representations under the symmetry. It is not possible to write the gauge invariant mass term for chiral fermions, and their masses are generated through the Yukawa coupling after the spontaneous breaking of the gauge symmetry. In case the gauge symmetry under which $\psi$ is chiral remains unbroken, then $\psi$ mass cannot be generated.  Chiral fermions induce anomalies which need to be canceled for the gauge theory to be unitary and renormalizable.  
\end{itemize}

\begin{figure}[ht]
\includegraphics[width=0.8\textwidth]{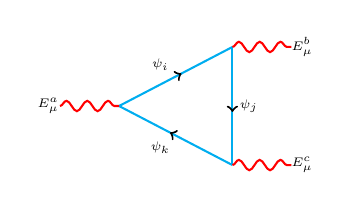}
\caption{The triangle diagram which can induce gauge anomalies. The chiral fermions $\psi_i$ run in the loop with gauge bosons $E^a_{\mu}$ emitted at each vertex. }
\label{TriangleD}
\end{figure}

Coming back to SM gauge symmetries, the quarks and leptons are chiral under both $SU(2)_L$ and $U(1)_Y$  gauge symmetries, and hence, the anomaly cancellation in the SM is not automatic. 
In four-dimensional chiral gauge theories like SM, three distinct anomalies have been identified. The first of these is a nonperturbative global anomaly, commonly referred to as the Witten $SU(2)$ anomaly \cite{Witten:1982fp}. This anomaly can occur in a $SU(2)$ gauge theory if it has an odd number of left-handed fermion doublets. However, this anomaly does not arise in the SM as it has an even number of $SU(2)_L$ fermion doublets.
Second is the perturbative triangular gauge anomaly encountered whenever a gauge theory has chiral fermions \cite{Adler:1969gk,Bardeen:1969md,Bell:1969ts}. In perturbation theory, gauge anomalies manifest within triangle loop diagrams featuring fermionic propagation and bosonic emission from vertices, illustrated in Fig.\ref{TriangleD}. 
%%%%%%%%%%
%%%%%%%%%%%%%
These diagrams should vanish to avoid the breakdown of gauge invariance and renormalizability of the theory. Note that vector fermions, do not contribute to these anomalies. However, since the SM has chiral fermions, one has to ensure that the gauge anomalies indeed cancel in SM. 
The third anomaly that we encounter is the mixed chiral gauge-gravitational anomaly \cite{Delbourgo:1972xb,Alvarez-Gaume:1983ihn}. This is also a perturbative anomaly, which must be canceled in order to ensure the general covariance of the theory. Again, its cancellation in SM is not automatic. 
The triangular anomaly cancellation conditions for all perturbative gauge (Eqs.~\eqref{anomaly1}-\eqref{anomaly2}) and mixed gauge-gravitational (Eq.~\eqref{anomalyG}) anomalies for SM symmetries are listed below
\begin{subequations}
\begin{align}
&[SU(3)_{C}]^2U(1)_{Y}=  \sum\limits_{i} Y_{Q^{i}} -  \sum\limits_{j} Y_{q^{j}} = 0\,, \label{anomaly1}
\\
&  [SU(2)_{L}]^2U(1)_{Y}=  \sum\limits_{i} Y_{L^{^{i}}} + 3\sum\limits_{j} Y_{Q^{^{j}}} = 0,\,
\\
&  [U(1)_{Y}]^3= \sum\limits_{i,j} ( Y_{L^{^{i}}}^{3} + 3 Y_{Q^{^{j}} }^{3}  ) - \sum\limits_{i,j} ( Y_{l^{^{i}}}^{3} + 3 Y_{q^{j}}^{3}  ) = 0 , \label{anomaly2}\,
\\
& [G]^2U(1)_{Y}= \sum\limits_{i,j} ( Y_{L^{^{i}}} + 3 Y_{Q^{^{j}}}  ) - \sum\limits_{i,j} ( Y_{l^{^{i}}} + 3 Y_{q^{j}} ) = 0.  \label{anomalyG}  
\end{align}
\end{subequations}
%%%%%%%%%%%%%%
Here, $Q = (u_\mathtt{L},d_\mathtt{L})^T$ and $L = (\nu_\mathtt{L},e_\mathtt{L})^T$ are quark and lepton doublets, respectively, while quark and lepton singlets are denoted by $q = \{u_\mathtt{R} ,d_\mathtt{R} \}$ and $l = e_\mathtt{R}$ respectively. 
The $Y_{\psi}$ represents the hypercharge of SM fermions, and $i,j =1,2,3$ are generational indices. Note that there are no equations corresponding to an odd number of $SU(3)_C$ or $SU(2)_L$ gauge bosons emitted from the triangular loop of Fig.~\ref{TriangleD} as the tracelessness condition of generators of these groups ensures that such diagrams vanish identically.  
In the following sections, we will discuss thoroughly the implications of anomaly cancellation requirements in SM and $U(1)_X$ gauge extensions of the SM.

%%%%%%%%%%%%%
\subsection{One generation of SM fermions}
\label{SubSec:Ano_one_gen_SM}
%%%%%%%%%%%%%%%

To begin with, let us first consider only one generation of SM fermions. In such a case, the hypercharges of SM fermions ($Y_\psi$) must add up in a way to cancel the anomalies. 
However, Eqs.~\eqref{anomaly1}-\eqref{anomalyG}  do not fix the hypercharge of SM fermions uniquely. These conditions give the following possible solutions for hypercharge \cite{Geng:1989tcu, Minahan:1989vd}:
\begin{itemize}
\item The first solution is $ Y_{u_{_{\mathtt{R}}}} + Y_{d_{_{\mathtt{R}}}} = 0$ and $Y_{L}=Y_{Q}=Y_{e_{_{\mathtt{R}}}}=0$.   This solution can be disregarded as no combination of weak isospin $T_{3}$ and hypercharge can produce correct electric charges of all fermions.
\item The second solution is the standard SM chiral hypercharge assignment, as shown in Table \ref{tab:SMCharge}. 
%%%%%%%%%%%%%%%%%%%%
\begin{table}[ht] 
 \renewcommand{\arraystretch}{1.5}
 \centering
 \begin{tabular}{|@{\hspace{7pt}} c  @{\hspace{5.5pt}}|@{\hspace{5.5pt}} c @{\hspace{5.5pt}}|@{\hspace{5.5pt}} c@{\hspace{5.5pt}}|@{\hspace{5.5pt}} c@{\hspace{5.5pt}}| @{\hspace{5.5pt}} c @{\hspace{5.5pt}}|@{\hspace{5.5pt}}c@{\hspace{7pt}}|}
 \hline 
$Q$ & $u_{_{\mathtt{R}}}$ & $d_{_{\mathtt{R}}}$ & $L$ & $e_{_{\mathtt{R}}}$& $\Phi$  \\ 
 \hline
 \hline
$\frac{Y}{3}$ & $\frac{4Y}{3}$ & $\frac{-2Y}{3}$ & $-Y$ & $-2Y$  &$Y$\\ 
 \hline
 \end{tabular}
 \caption{SM particles hypercharge assignment, the normalization can be fixed by defining the relationship between hypercharge and electric charge and then requiring that the electric charge for the neutrino is zero. The two often used choices are $Y=1$ or $Y = 1/2$.}
 \label{tab:SMCharge} 
\end{table}
%%%%%%%%%%%%%%%%%%

\item One more solution can be found by interchanging the hypercharges of $u_{_{\mathtt{R}}}$ and $d_{_{\mathtt{R}}}$ i.e. $Y_{u_{_{\mathtt{R}}}}=\frac{-2Y}{3}$ and $Y_{d_{_{\mathtt{R}}}}=\frac{4Y}{3}$ in Table~\ref{tab:SMCharge}, but again this leads to incorrect electric charges of fermions.
\end{itemize}
Hence, the gauge anomaly alone does not uniquely determine the hypercharge assignment of fermions.
However, the additional requirement that the hypercharge assignment must result in the correct electric charges for SM fermions leads to the unique assignment shown in Table~\ref{tab:SMCharge}. 
%%%%%%%%%%%%%%%%%%

Another way to fix hypercharges uniquely is to take into consideration the mass generation mechanism of these fermions. In SM the fermions mass is generated via Yukawa interactions,   
%%%%%%%%%%%%%%%%%%%%%%%%%%%%%%%%%%%%%%%%%%%%%%%%%%%%%%%%%%%%%%%%%%%%%%%%%
\begin{equation}
\label{yukawa terms of e u d SM}
-\mathscr{L}_{yukawa} \supset Y_{e}\overline{L} \Phi e_{\mathtt{R}} +Y_{u} \overline{Q} \tilde{\Phi} u_{\mathtt{R}}  + Y_{d}\overline{Q} \Phi d_{\mathtt{R}} + \text{h.c.}\,. 
\end{equation} 
%%%%%%%%%%%%%%%%%%%
Here, $\Phi$ denotes the SM Higgs doublet, and $\tilde{\Phi}$ corresponds to $i\sigma_{2}\Phi^{*}$, where $\sigma_{2}$ is the second Pauli matrix. The invariance of the Yukawa couplings under hypercharge symmetry gives us the two extra conditions,
\begin{equation}\label{yzSM}
Y_{u_{_{\mathtt{R}}}}= Y_{Q} + Y_{L} - Y_{e_{_{\mathtt{R}}}}, ~ \text{and}~Y_{d_{_{\mathtt{R}}}} = Y_{Q} - Y_{L} + Y_{e_{_{\mathtt{R}}}} .
\end{equation}
%%%%%%%%%%%%%%%%%%%%%
These extra conditions, along with the anomaly cancellation conditions of \eqref{anomaly1}-\eqref{anomalyG}, also fix the SM hypercharge uniquely to the one listed in Table \ref{tab:SMCharge}.

In summary, for one generation of SM fermions, the hypercharges cannot be uniquely fixed by anomaly cancellation requirements alone. In addition one has to also demand that the hypercharge assignment leads to correct electric charges of the SM fermions and/or leads to generation of fermions masses through their Yukawa couplings with the SM Higgs boson.

%%%%%%%%%%%%%%%%%%%%%%%%%%%%%%%%%%%%%%%%%%%%%%%%%%%%%%%%%%%%%%%%%%%%%%%%%%%%%%%
\subsection{Three generations of SM fermions }
\label{SubSec:Ano_three_gen_SM}
%%%%%%%%%%%%%%%%%%%%%%%%%%%%%%%%%%%%%%%%%%%%%%%%%%%%%%%%%%%%%%%%%%%%%%%%%%%%%%%

For three generations of SM fermions, apart from the previously mentioned solutions, one can have additional solutions to Eqs.~\eqref{anomaly1}-\eqref{anomalyG}. For example: 
\begin{itemize}
\item One such solution is obtained when hypercharges of all leptons are zero and the hypercharges of the three generations of quark doublets are  $Y_{Q^{i}} = - Y_{Q^{j}} = Y, \, \, Y_{Q^{k}} = 0$ where $i,j,k = 1,2,3~ \&~ i \neq j \neq k$. In addition the hypercharges of quark singlets should be such that $Y_{u_{_{\mathtt{R}}}^{l}} = - Y_{u_{_{\mathtt{R}}}^{m}} = Y^{'}, \, \, Y_{u_{_{\mathtt{R}}}^{n}} = 0$ where $l,m,n = 1,2,3~ \&~ l \neq m \neq n$ and $Y_{d_{_{\mathtt{R}}}^{r}} = - Y_{d_{_{\mathtt{R}}}^{s}} = Y^{''}, \, \, Y_{d_{_{\mathtt{R}}}^{t}} = 0$ where $r,s,t = 1,2,3~ \&~ r \neq s \neq t$. 
Clearly, like before, this kind of hypercharge assignment can be rejected as the correct electric charges of the quarks and leptons cannot be obtained.

\item One specific case of the above solution arises when, $i=l=r, \, \, j=m=s,$ and $k=n=t$. In this particular case, the hypercharge assignment is consistent with the gauge anomalies cancellation as well as mass generation, but still, the correct electric charges cannot be obtained.  Furthermore, one will not be able to generate the correct quark mixing structure without the addition of BSM particles. 

\item Another anomaly-free solution can be obtained by making all quark hypercharges to be zero and $Y_{L^{i}} = - Y_{L^{j}} = Y, \, \, Y_{L^{k}} = 0$ where $i,j,k = 1,2,3~ \&~ i \neq j \neq k$. In addition, the hypercharges of leptons singlets should be such that $Y_{e_{_{\mathtt{R}}}^{l}} = - Y_{e_{_{\mathtt{R}}}^{m}} = Y', \, \, Y_{e_{_{\mathtt{R}}}^{n}} = 0$ where $l,m,n = 1,2,3~ \&~ l \neq m \neq n$. Again like previous cases, correct electric charges cannot be obtained.

\end{itemize}

In summary, for the full three generations of SM fermions, the anomaly cancellation conditions \eqref{anomaly1}-\eqref{anomalyG} gives several possible solutions, some of which can even lead to correct mass generation (but not mixing) for the fermions. However, none of these extra solutions can be used to correctly obtain the electric charges of all SM fermions. Therefore, the standard SM hypercharge assignment remains unique even with three generations of SM fermions. 

It should be noted that some of the solutions discussed here, particularly in this subsection, can be valid solutions in BSM models with new $U(1)_{X}$ symmetries. This is because, for a BSM $U(1)_{X}$ symmetry, there is no necessity to relate the $U(1)_{X}$ charges of fermions with their electric charges. Thus, the charges of SM fermions under $U(1)_{X}$ get constraints only from gauge anomaly and their mass generation mechanism, both of which some solutions discussed here satisfy. Hence, one can explore these symmetries in BSM scenarios and recover the correct structure of CKM by adding vectorlike fermions in a way similar to Ref.~\cite{Bonilla:2017lsq}. To the best of our knowledge, such models except $L_i - L_j$ \cite{He:1990pn}, have not been explored in the literature. 
In the following section, we will discuss in detail the new $U(1)_X$ gauge symmetries in BSM scenarios.

%%%%%%%%%%%%%%%%%%%%%%%%%%%%%%%%%%%%%%%%%%%%%%%%%%%%%%%%%%%%%%%%%%%%%%%%%%%%%%%%%%
\section{anomaly cancellation for $U(1)_{X}$ gauge symmetry}\label{sec:u1x-anomaly} 

%%%%%%%%%%%%%%%%%%%%%%%%%%%%%%%%%%%%%%%%%%%%%%%%%%%%%%%%%%%%%%%%%%%%%%%%%%%%%%%%%%%%%%%%%

Having discussed the anomaly cancellation in SM, let us now look at gauge anomaly conditions in the presence of a new  $U(1)_{X}$ gauge symmetry. The new $U(1)_{X}$ gauge symmetry leads to additional anomalies which need to be canceled. In most of the cases, the anomaly induced by SM fermions needs to be canceled by some BSM fermions. In this case, along with (Eqs.~\eqref{anomaly1}-\eqref{anomalyG}), one must satisfy six additional anomaly cancellation conditions given by the following:
\begin{subequations}
\label{U1x anomaly cancellation}
\begin{align}
&[SU(3)_{C}]^2[U(1)_{X}]=  \sum\limits_{i} X_{Q'^{^{i}}} -  \sum\limits_{j} X_{q'^{^{j}}} = 0.\label{ano1} 
\\& [SU(2)_{\mathtt{L}}]^2[U(1)_{X}]=  \sum\limits_{i} X_{L'^{^{i}}} + 3\sum\limits_{j} X_{Q'^{^{j}}} = 0. \label{ano2}
\\& [U(1)_{Y}]^2 [U(1)_{X}] = \sum\limits_{i,j} ( Y_{L'^{^{i}}}^2   X_{L'^{^{i}}} + 3 Y_{Q'^{^{j}}}^2   X_{Q'^{^{j}}}  )  - \sum\limits_{i,j} ( Y_{l'^{^{i}}}^2 X_{l'^{^{i}}} + 3 Y_{q'^{^{j}}}^2 X_{q'^{^{j}}}  ) = 0. \label{ano3}
\\& [U(1)_{Y}] [U(1)_{X}]^2 = \sum\limits_{i,j} ( Y_{L'^{^{i}}}   X_{L'^{^{i}}}^2 + 3 Y_{Q'^{^{j}}}   X_{Q'^{^{j}}}^2  )  - \sum\limits_{i,j} ( Y_{l'^{^{i}}} X_{l'^{^{i}}}^2 + 3 Y_{q'^{^{j}}} X_{q'^{^{j}}}^2  ) = 0. \label{ano4}
\\&  [U(1)_{X}]^3= \sum\limits_{i,j} ( X_{L'^{^{i}}}^{3} + 3 X_{Q'^{^{j}}}^{3}  ) - \sum\limits_{i,j} ( X_{l'^{^{i}}}^{3} + 3 X_{q'^{^{j}}}^{3}  ) = 0. \label{ano5}
\\& [G]^2[U(1)_{X}]= \sum\limits_{i,j} ( X_{L'^{^{i}}} + 3 X_{Q'^{^{j}}}  ) - \sum\limits_{i,j} ( X_{l'^{^{i}}} + 3 X_{q'^{^{j}}} ) = 0. \label{ano6}
\end{align}
\end{subequations}
%%%
%
Here $X_{\psi}$ is the $U(1)_{X}$ charge of the fermions, $L'=\{L, \Psi \} $ denotes all SM ($L$) and BSM ($\Psi$) lepton $SU(2)_{L}$ doublets, while $l'=\{l, \psi \}$ are SM $(l)$ and BSM $(\psi)$ lepton $SU(2)_{L}$ singlets. 
Similarly, $Q'=\{ Q, \Psi_{c} \}$ denotes all SM $(Q)$ and BSM ($\Psi_{c}$) $SU(2)_{L}$ doublet fermions charged under $SU(3)_{C}$ and $q'=\{q,\psi_{c}\}$ denotes all SM ($q$) and BSM ($\psi_{c}$) $SU(2)_{L}$ singlet fermions charged under $SU(3)_{C}$.

In writing Eqs.(\ref{ano1}-\ref{ano6}), we have assumed that any new chiral BSM fermion is either in trivial or in fundamental representations of non-Abelian SM gauge groups, i.e. $SU(3)_{C}$ and  $SU(2)_L$.
 Obviously, any BSM fermion which is vector under the SM, as well as the $U(1)_X$ gauge symmetries, will not play any role in anomaly cancellation. In the subsections below, we discuss some of the known solutions which lead to anomaly-free $U(1)_X$ gauge symmetries for the case of both vector and chiral BSM fermions.

%%%%%%%%%%%%%%%%%%%%%%%%%%%%%%%%%%%%%%%%%%%%%%%%%%%%%%%%%%%%%%%%%%%%%%%%%%%%%%%%%%%%%
\subsection{Vector solutions for $U(1)_{X}$ }
\label{SubSec:U(1)x_vector_solu}
%%%%%%%%%%%%%%%%%%%%%%%%%%%%%%%%%%%%%%%%%%%%%%%%%%%%%%%%%%%%%%%%%%%%%%%%%%%%%%%%%%%%%
In this section, we discuss the ``vector solutions" to $SU(3)_C \otimes SU(2)_L \otimes U(1)_Y \otimes U(1)_X$ gauge anomaly cancellation conditions. These solutions are characterized by vector charges of SM fermions under $U(1)_{X}$. 
Depending on the $U(1)_X$ charges of the SM fermions, we have two types of solutions to the gauge anomalies. The first type of solution does not require the addition of any BSM fermions, as the anomalies can be canceled by appropriately choosing the $U(1)_X$ charge of SM fermions themselves. The second type of solution requires the addition of BSM fermions, whose $U(1)_{X}$ charges are the same as charges of one or another SM fermions. Typically the popular choice is to introduce SM gauge singlet fermions having the same $U(1)_X$ charge as SM neutrinos. 

First, let us discuss the solution that does not require the addition of any BSM fermion.
\begin{itemize}
    \item \underline{$\mathbf{B_{i}-B_{j}}$}:  This solution to anomaly cancellation is obtained when:
\begin{eqnarray}
&& X_{Q^{i}} = - X_{Q^{j}} = X, \, \, X_{Q^{k}} = 0; \qquad   i,j,k = 1,2,3~ \,\& \,~ i \neq j \neq k \nonumber \\
&& X_{u_{_{\mathtt{R}}}^{l}} = - X_{u_{_{\mathtt{R}}}^{m}} = X, \, \, X_{u_{_{\mathtt{R}}}^{n}} = 0; \qquad  l,m,n = 1,2,3~ \, \&\,~ l \neq m \neq n \nonumber \\
&&  X_{d_{_{\mathtt{R}}}^{r}} = - X_{d_{_{\mathtt{R}}}^{s}} = X, \, \, X_{d_{_{\mathtt{R}}}^{t}} = 0; \qquad  r,s,t = 1,2,3~ \, \&~ r \neq s \neq t \nonumber \\
&& X_{L^{i}}=X_{l^{^{j}}} = 0; \quad   i,j= 1,2,3~~ \forall~ i,j. \nonumber
\end{eqnarray}
  Note that a simplifying yet viable choice will be $i=l=r, \, \, j=m=s,$ and $k=n=t$  in which case the symmetry can be referred to as $B_i - B_j$. 
This is a flavor-dependent gauge symmetry. Under this symmetry, anomalies induced in Eqs.\eqref{ano2} and \eqref{ano3} by one generation of quarks ($Q^i, q^i$) are canceled by another generation ($Q^j, q^j$). Other anomaly equations are also satisfied due to the vectorial nature of this symmetry i.e. $X_{Q^i} = X_{q^i}$ under $U(1)_{B_i - B_j}$. The mass of all the fermions can be generated via Yukawa terms, with SM Higgs uncharged under $U(1)_{B_i - B_j}$. However, these charge assignments do not lead to correct quark mixing. Hence, new BSM quarks that are vector under both SM and $U(1)_{X}$ gauge symmetries are needed to achieve the correct structure of the CKM matrix. To the best of our knowledge, these solutions have not been explored in the literature, and we are the first ones to point them out.

\item \underline{$\mathbf{L_{i}-L_{j}}$}: This anomaly cancellation solution can be written as follows:
\begin{eqnarray}
&& X_{L^{i}} = -X_{L^{^{j}}}=X ,\,\, X_{L^{k}}=0; \qquad  i,j,k=1,2,3 ~ \,\& \, ~ i \neq j \neq k \nonumber \\
&& X_{e_{_{\mathtt{R}}}^{l}} = - X_{e_{_{\mathtt{R}}}^{m}} = X, \, \, X_{e_{_{\mathtt{R}}}^{n}} = 0 ; \qquad l,m,n = 1,2,3~ \, \& \,~ l \neq m \neq n \nonumber \\
&& X_{Q^{i}}=X_{q^{^{j}}} = 0; \quad   i,j= 1,2,3~~ \forall~ i,j. \nonumber
\end{eqnarray}  
    
 Again,  if $i=l, \, \, j=m,$ and $k=n$ then this symmetry reduces to the $L_{i}-L_{j}$ gauge symmetry~\cite{He:1990pn}. This is again a flavor dependent symmetry acting only on the lepton sector. 
 Like the previous case, here too the symmetry allows for the generation of charged lepton masses via Yukawa couplings with SM Higgs, but neutrinos still remain massless. Again, correct neutrino masses and mixing can be achieved by adding BSM fermions, which are vector under both $U(1)_{X}$ and SM gauge symmetries.            
\end{itemize}
%%%%%%%%%%%%%%%%%%%%%%%%%%%%%%%%%%%%%%%%%%%%%%%%%%%%%%%%%%%%%%%%%%%%%%%%%%%%%%%%%%%%%%%%%%%%%%%%%

We now focus our attention on solutions that require BSM fermions to cancel gauge anomalies.
These symmetries are widely studied in the literature, with various different solutions known to exist~\cite{Allanach:2018vjg,Ghosh:2024cxi}. 
It is important to note that any BSM fermion pair ($f_{\mathtt{L}},f_{\mathtt{R}}$)  vector under both SM and $U(1)_{X}$ symmetries cannot help in anomaly cancellation. Similarly, any BSM fermion pair which is singlet under SM gauge symmetries and vector under $U(1)_{X}$ also does not help in anomaly cancellation.
Finally, since the $U(1)_X$ charge of the SM fermions can be chosen in many ways, the BSM fermion charges in each case should be chosen appropriately so that they can cancel the induced anomalies.  
 Some of the well-known solutions where the SM singlet fermion (taken to be right handed) carrying the same $U(1)_{X}$ charge as SM lepton doublet are introduced to cancel anomalies are presented below.

%%%%%%%%%%%%%%%%%%%%%%%%%%%%%%%%%%%%%%%%%%%%%%%%%%%%%%%%%%%%%%%%%%%%%%%%%%%%%%%%%

\begin{itemize}
\item \underline{$\mathbf{B-L} $}: Under this symmetry charges of fermions are given as 
%%%%%%%%%%%%
\begin{eqnarray}
&& X_{Q^{i}} = X_{q^{j}} = 1/3; \quad   i,j=1,2,3\,,  \nonumber \\
&& X_{L^{i}}=X_{l^{^{j}}}=X_{f^{i}} = -1;\quad \forall~i,j \nonumber.
\end{eqnarray}  
%%%%%%%%%%%%%%%%%%%%%%%%%%%%%%%%%%%%%%%%%%%%%%%%%%%%%%%%%%%%%%%%%%%%%%%%%%%%%
This is one of the most popular flavor-independent symmetries. Under this symmetry, the anomaly induced by the last two equations Eqs. (\ref{ano5}-\ref{ano6}) are canceled by introducing three SM singlet right-handed fermions ($f_{i}$) with charge $X_{f^{i}}=-1$. Equation \eqref{ano1} is automatically satisfied due to vectorial nature of quarks under $U(1)_{X}$. Anomaly induced by quarks in Eq.\eqref{ano2} should be canceled by leptons, and that is the reason for the relative sign between $B$ and $L$ and a factor of 3 in charges of quarks under this symmetry. Equations (\ref{ano3}-\ref{ano4}) does not induce any anomaly due to the vectorial nature of both quarks and leptons and the structure of $B-L$.  
%%%%%%%%%%%%%%%%%%%%%%%%%%%%%%%%%%%%%%%%%%%%%%%%%%%%%%%%%%%%%%%%%%%%%%%%%%%%
\item \underline{$\mathbf{B-3L_{i}}$}: These are flavor-dependent symmetry, where all generations of quarks have the same $U(1)_{X}$ charge, however, only one generation of SM leptons carry the $U(1)_{X}$ charge \cite{Ma:1997nq,Lee:2010hf}. 
Charges of fermions under this symmetry are given as,
%%%%%%%%%%%%%%%%%%%%%%%%%%%%%%%%%%%%%%%%%%%%%%%%%%%%%%%%%%%%%%%%%%%%%%%%%%%%
\begin{eqnarray}
&& X_{Q^{i}} = X_{q^{j}} = 1/3; \quad   i,j=1,2,3\,,  \nonumber \\
&& X_{L^{i}}=X_{l^{i}}= X_{f} = -3,\, X_{L^{k}}=X_{l^{k}} =0 ; \quad  i,k=1,2,3 ~ \,\& \, ~ i \neq k  \nonumber  \nonumber.
\end{eqnarray}  
%%%%%%%%%%%%%%%%%%%%%%%%%%%%%%%%%%%%%%%%%%%%%%%%%%%%%%%%%%%%%%%%%%%%%%%%%%%%%

Under this symmetry, Eq.\eqref{ano1} and Eq.\eqref{ano4} do not induce any anomaly due to the vectorial nature of symmetry. Compared to the $B-L$ case, a factor of 3 in $L_i$ charge is required to cancel anomalies induced by Eq.\eqref{ano2} and Eq.\eqref{ano3}, as only one generation of leptons is charged under $U(1)_{X}$. To cancel anomalies induced by Eq.\eqref{ano5} and Eq.\eqref{ano6}, one SM singlet right-handed fermion with charge $X_{f}=-3$ is introduced.   

\item \underline{$\mathbf{B_{i}-3L_{j}}$}: Here, the flavor dependency arises from both the quark and lepton sectors. Charges of fermions under this symmetry are given as \cite{Bonilla:2017lsq,Allanach:2022blr,Alonso:2017uky,DeRomeri:2023ytt},
%%%%%%%%%%%%%%%%%%%%%%%%%%%%%%%%%%%%%%%%%%%%%%%%%%%%%%%%%%%%%%%%%%%%%%%%%%%%
\begin{eqnarray}
&& X_{Q^{i}}=X_{q^{i}} =1,\,X_{Q^{j}}=X_{q^{j}}=0; \quad  i,j=1,2,3 ~ \,\& \, ~ i \neq j\,,  \nonumber  \\
&& X_{L^{^{j}}}=X_{l^{^{j}}} = X_{f} = -3,\, X_{L^{^{k}}}=X_{l^{^{k}}} =0;\quad  j,k=1,2,3 ~ \,\& \, ~ j \neq k  \nonumber.
\end{eqnarray}  
%%%%%%%%%%%%%%%%%%%%%%%%%%%%%%%%%%%%%%%%%%%%%%%%%%%%%%%%%%%%%%%%%%%%%%%%%%%%%
%
The anomaly cancellation is similar to the $B - 3L_i$ scenario, with the key difference being that only one generation of quarks is charged. However, since the charges are still vectorlike, this does not create any issues, and the anomalies cancel out as in the previous case with the introduction of one SM singlet right-handed fermion with charge $X_{f}=-3$.
%%%%%%%%%%%%%%%%%%%%%%%%%%%%%%%%%%%%%%%%%%%%%%%%%%%%%%%%%%%%%%%%%%%%%%%%%%%%%
\item Some other solutions $B-2L_{i}-L_{j}$, $B-\frac{3}{2}(L_{i}+L_{j})$, and $B_{1}-yB_{2}+(y-3)B_{3}+L_{i}+L_{j}$ \cite{Farzan:2015doa,Allanach:2018vjg, DeRomeri:2024dbv,AtzoriCorona:2022moj}, for fixed $i$ and $j$, are also explored in the literature.
 
\end{itemize}

%%%%%%%%%%%%%%%%%%%%%%%%%%%%%%%%%%%%%%%%%%%%%%%%%%%%%%%%%%%%%%%%%%%%%%%%%%%%%%%%%%
\subsection{Chiral solutions for $U(1)_{X}$}
\label{SubSec:U(1)x_Chiral_solu}
%%%%%%%%%%%%%%%%%%%%%%%%%%%%%%%%%%%%%%%%%%%%%%%%%%%%%%%%%%%%%%%%%%%%%%%%%%%%%%%%%%

Now let us look at the possible ``chiral solutions," i.e. the solutions to anomaly cancellation where the SM or BSM fermions are chiral under the $U(1)_X$ symmetry.
The chiral nature of fermion charges complicates the anomaly cancellation conditions.
Furthermore, generating the fermion masses and SM fermion mixings can also be challenging, as in this case, the $U(1)_{X}$ symmetry can forbid some of the Yukawa couplings.
Therefore, nontrivial chiral solutions that satisfy both these requirements are relatively rare.
In the literature, some attempts have been made in this direction, and some of the known solutions are presented below.   

\begin{itemize}
\item  \underline{\bf{Chiral} $\mathbf{B-L}$}: This is an alternate solution to the $B-L$ gauge symmetry where the BSM fermions ($f_i$) are chiral under the symmetry~\cite{Montero:2007cd,Ma:2014qra,Ma:2015raa,Ma:2015mjd}.  
 Under this symmetry, charges of fermions are given as
%%%
\begin{eqnarray}
&& X_{Q^{i}} = X_{q^{j}} = 1/3; \quad   i,j=1,2,3\,,  \nonumber \\
&& X_{L^{i}}=X_{l^{^{j}}} = -1;\quad \forall~i,j\,, \nonumber \\
&& X_{f^{1}}=-4,\,X_{f^{2}}=-4,\,X_{f^{3}}=5.
\end{eqnarray}  
%%%%%%%%%%%%%%%%%%%%%%%%%%%%%%%%%%%%%%%%%%%%%%%%%%%%%%%%%%%%%%%%%%%%%%%%%%%%%    
Again, Eqs. \eqref{ano1}-\eqref{ano4} are anomaly-free, as described previously. The anomaly induced by the last two equations needs to be canceled by BSM fermions. A chiral solution, where $U(1)_{B-L}$ charge of fermions (taken to be right handed) are  $(-4,-4,5)$ can also cancel these anomalies.

\item  \underline{\bf{ New Chiral} $\mathbf{B-L}$}:
Additionally, we have identified a new solution where charges of SM gauge singlet BSM fermions (taken to be right-handed) $f_i$; $i =1,2,3$ under $U(1)_{B-L}$ are $(6,\frac{-17}{3},\frac{-10}{3})$.
Charges of SM fermion under $U(1)_{B-L}$ remain same as before. 

 \item \underline{\bf{ Gauged $\mathbf{U(1)_B}$ }}:  
 It is well known that the accidental global $U(1)_B$ symmetry of SM  is anomalous symmetry. 
This symmetry can be made anomaly-free by adding fermions appropriately \cite{Rahul:2024N}. Charges of SM fermions under this symmetry can be written as 
\begin{eqnarray}
&& X_{Q^{i}} = X_{q^{j}} = X; \quad   i,j=1,2,3\,,  \nonumber \\
&& X_{L^{i}}=X_{l^{^{j}}} = 0;\quad \forall~i,j \nonumber 
\end{eqnarray}  
The anomaly is only induced by Eq.\eqref{ano2} and Eq.\eqref{ano3}, other equations are all zero due to the vectorial nature of $U(1)_{B}$. The anomaly induced by Eq.(\ref{ano2}-\ref{ano3}) could be eliminated by adding BSM fermions that are non-trivially charged under $SU(2)_{L}$ and hypercharge.

 \item  \underline{\bf{Gauged $\mathbf{U(1)_L}$}}: It is another accidental global $U(1)$ symmetry of SM. Charges of SM fermions under this symmetry are given as
\begin{eqnarray}
&& X_{L^{i}} = X_{l^{^{j}}} = X; \quad   i,j=1,2,3\,,  \nonumber \\
&& X_{Q^{i}}=X_{q^{^{j}}} = 0;\quad \forall~i,j. \nonumber 
\end{eqnarray}  
%%%%%%%%%%%%%%%%%%%%%%%%%%%%%%%%%%%%%%%%%%%%%%%%%%%%%%%%%%%%%%%%%%%%%%%%%%%%% 
In this case, only Eqs.\eqref{ano1} and Eq.\eqref{ano4} are anomaly-free; all other equations induce anomalies. Anomalies of the last two equations can be canceled by adding SM singlet three vectors like right-handed fermions with $U(1)_{X}$ charge $X_{f^{i}}=-X$. Additionally, to cancel anomalies introduced by Eq.(\ref{ano2}-\ref{ano3}) we need BSM fermions that are nontrivially charged under $SU(2)_{L}$ and hypercharge \cite{Rahul:2024N}.

\end{itemize}

An important thing to note is that most of these solutions are limited in their chirality, i.e., either only a small subset of SM fermions carry $U(1)_X$ charge or only the new BSM fermions are chiral, while SM fermions are kept vector under the $U(1)_X$ symmetry. 
This is unlike the SM hypercharge under which all SM fermions are chiral.  
Solutions where both SM and BSM fermions are chiral in nature are not well explored. In literature, certain attempts have been made in this direction, considering a linear combination of hypercharge and $B-L$ charge \cite{Das:2016zue} or a $U(1)_R$ symmetry where only right-handed fermions are charged \cite{Jana:2019mez}.
One may wonder whether truly chiral $U(1)_X$ solutions are possible or not?
The answer is yes. Indeed, such solutions are possible and a few are discussed in \cite{Appelquist:2002mw}.
However, as we show, a whole class of yet unexplored solutions can be found. Furthermore, they can be potentially related to the existence of dark matter, with the BSM fermions needed for anomaly cancellation belonging to the dark sector.  
The lightest of the dark fermions can then be a good dark matter candidate with the $U(1)_X$ gauge boson $Z'$ being the mediator connecting the dark sector to the visible sector. 
For this reason we call this new class of $U(1)_X$ symmetries as dark hypercharge (DHC) symmetries.
In the next section, we explore such truly chiral solutions.

%%%%%%%%%%%%%%%%%%%%%%%%%%%%%%%%%%%%%%%%%%%%%%%%%%%%%%%%%%%%%%%%%%%%%%%%%%%%%%%%%%
\section{Dark Hypercharge Solutions} \label{Sec_DHC}
\label{Sec:U(1)x_DarkHypercharge_solu}
%%%%%%%%%%%%%%%%%%%%%%%%%%%%%%%%%%%%%%%%%%%%%%%%%%%%%%%%%%%%%%%%%%%%%%%%%%%%%%%%%%
 
In this section, we explore chiral solutions to $SU(3)_C \otimes SU(2)_L \otimes U(1)_Y \otimes U(1)_X$ gauge anomaly cancellation conditions. We consider SM fermions to be chiral under DHC symmetry. To cancel the gauge anomalies induced by these chiral SM fermions, we introduce new SM singlet right-handed dark fermions (DFs). Let's look at anomaly cancellation conditions with this setup. In the most general case, every generation of fermions has different DHC charges $X_{\psi'}$. Here, $\psi'=\{L^{i},Q^{i},e_{\mathtt{R}}^{i},u_{\mathtt{R}}^{i},d_{\mathtt{R}}^{i}, f^{k}\}$ and $i= 1,2,3$ is a flavor index. New BSM fermions are denoted by $f^k$, while index $k$ represents the generations and could take any integral value.  Putting these charges in Eqs.(\ref{ano1}-\ref{ano6}), we get
%%%%%%%%%%%%%%%%%%%%%%%%%%%%%%%%%%%%%%%%%%%%%%%%%%%%%%%%%%%%%%%%%%%%%%%%%%%%%%%%%%%%%%%%%%%%
\begin{subequations}\label{fac}
\begin{align}
&[SU(3)_{C}]^2[U(1)_{X}]= \sum_{i=1}^{3}(2 X_{Q^{^{i}}} -  X_{u_{_{\mathtt{R}}}^{^{i}}}-X_{d_{_{\mathtt{R}}}^{^{i}}}) \label{Eq:anoUx1} = 0\,, 
\\&[SU(2)_{\mathtt{L}}]^2[U(1)_{X}]= \sum_{i=1}^{3}( X_{L^{^{i}}} +3 X_{Q^{^{i}}}) = 0\label{Eq:anoUx2}\,,
\\&[U(1)_{Y}]^2 [U(1)_{X}]= \sum_{i=1}^{3} ( X_{L^{^{i}}} + \frac{1}{3}  X_{Q^{^{i}}} -2  X_{e_{_{\mathtt{R}}}^{^{i}}} -\frac{8}{3}X_{u_{_{\mathtt{R}}}^{^{i}}}-\frac{2}{3}  X_{d_{_{\mathtt{R}}}^{^{i}}} ) = 0\label{Eq:anoUx3}\,,  
\\&[U(1)_{Y}] [U(1)_{X}]^2=  \sum_{i=1}^{3} \bigl\{   (X_{Q^{^{i}}})^{2}-(X_{L^{^{i}}})^{2}  + (X_{e_{_{\mathtt{R}}}^{^{i}}})^2 -2 (X_{u_{_{\mathtt{R}}}^{^{i}}})^2 + (X_{d_{_{\mathtt{R}}}^{^{i}}})^2  \bigl\} = 0 \label{Eq:anoUx4} \,,
\\ & [U(1)_{X}]^3= \sum_{i=1}^{3} \left[ 2(X_{L^{^{i}}})^{3}  +6 (X_{Q^{^{i}}})^{3}   - (X_{e_{_{\mathtt{R}}}^{^{i}}})^{3} -3 \bigl\{ (X_{u_{_{\mathtt{R}}}^{^{i}}})^{3}  + (X_{d_{_{\mathtt{R}}}^{^{i}}})^{3} \bigl\} \right]  - \sum_{k=1}^{m} (X_{f^{^{k}}})^{3} = 0 \label{Eq:anoUx5} \,,
\\&[G]^2[U(1)_{X}]= \sum_{i=1}^{3} \bigl\{ 2X_{L^{^{i}}}  + 6 X_{Q^{^{i}}} -X_{e_{_{\mathtt{R}}}^{^{i}}}  -3\left( X_{u_{_{\mathtt{R}}}^{^{i}}} + X_{d_{_{\mathtt{R}}}^{^{i}}}\right) \bigl\}  - \sum_{k=1}^{m}X_{f^{^{k}}} = 0 \label{Eq:anoUx6}\,. 
\end{align}
\end{subequations}
%%%%%%%%%%%%%%%%%%%%%%%%%%%%%%%%%%%%%%%%%%%%%%%%%%%%%%%%%%%%%%%%%%%%%%%%%%%%%%%%%%%%%%%%%%%
We study the ``dark hypercharge" solutions of Eqs.(\ref{Eq:anoUx1}-\ref{Eq:anoUx6}) under three different scenarios :
\begin{itemize}
     \item \textbf{ One generation scenario (SI) :} Only one generation of SM fermions are charged under DHC ($X_{\psi^{i}} \neq 0, 
     X_{\psi^{j}} = X_{\psi^{k}} = 0, \quad i,j,k=1,2,3~\&~ j,k \neq i $).
    \item \textbf{ Two generation scenario (SII) :} Two generations of SM fermions of each type share the same charge under DHC, while one generation remains uncharged ($X_{\psi^{i}}= X_{\psi^{j}}$, $X_{\psi^{k}}=0$,$\quad i,j,k=1,2,3~\&~i,j \neq k$).
     \item \textbf{ Three generation scenario (SIII) :} All three generations of SM fermions are charged under the new symmetry, and charges of a given fermion type are identical across generations ($X_{\psi^{i}} = X_{\psi^{j}}= X_{\psi^{k}},\quad i,j,k=1,2,3$).
\end{itemize}   
%%%%%%%%%%%%%%%%%%%%%%%%%%%%%%%%%%%%%%%%%%%%%%%%%%%%%%%%%%%%%%%%%%%%%%%%
The chiral nature of the SM fermions under DHC introduces additional complications in their mass generation mechanism. This is because chiral charge assignments generally forbids the SM Yukawa couplings,    
%%%%%%%%%
\begin{equation}
\label{Eq:SM_Yukawa}
-\mathscr{L}_{yukawa} \supset Y_{e}^{ij}\overline{L^{i}} \Phi e_{{\mathtt{R}}}^{j} +Y_{u}^{ij} \overline{Q^{i}} \tilde{\Phi} u_{{\mathtt{R}}^{i}}  + Y_{d}^{ij} \overline{Q^{i}} \Phi d_{{\mathtt{R}}}^{j} + \text{h.c.}\,. 
\end{equation} 
%%%
 In order to generate mass of SM fermions\footnote{Alternatively, SM fermion masses and mixing could be generated by adding new Higgs doublets.}, we also demand the invariance of SM Yukawas under DHC symmetry.
%%%
In all three above-mentioned scenarios, the constraints from Yukawas to generate masses of SM fermions can be written in terms of Higgs DHC charge $X_{\Phi}$ as,
\begin{equation} \label{Eq:Higgs_Charge_Final}
X_{_{\Phi}}=X_{L}-X_{e_{_{\mathtt{R}}}}=X_{Q}-X_{d_{_{\mathtt{R}}}}=X_{u_{_{\mathtt{R}}}}-X_{Q}.
\end{equation}
%%%%
It should be noted that, in the first two scenarios (SI and SII), all generations of SM fermions do not have the same DHC charge. Therefore, for these cases Eq.\eqref{Eq:Higgs_Charge_Final} should be understood to be applicable only to the generations charged under DHC symmetry. 
Furthermore, in both of these scenarios even satisfying Eq.\eqref{Eq:Higgs_Charge_Final} is not enough to generate the masses of all SM fermions. Consequently, in these scenarios, the inclusion of at least one more Higgs doublet is required alongside the SM Higgs\footnote{For the sake of brevity, we will call $\Phi$ as SM Higgs, even though it carries a nontrivial $U(1)_X$ charge. } to generate masses and mixings of the SM fermions. In the third scenario (SIII), the SM Yukawa structure (see Eq. \eqref{Eq:SM_Yukawa}) remains preserved, and masses as well as mixings of all SM fermions are generated by single Higgs doublet. 
%%%
Using Eq.\eqref{Eq:Higgs_Charge_Final} we can write $X_{u_{_{\mathtt{R}}}}$ and $X_{d_{_{\mathtt{R}}}}$ as,
\begin{equation}\label{Eq:quark_singlet_charges_to_satisfy_Yukawa}
X_{u_{_{\mathtt{R}}}} = X_{Q} + X_{L} - X_{e_{_{\mathtt{R}}}}, ~ \text{and}~X_{d_{_{\mathtt{R}}}} = X_{Q} - X_{L} + X_{e_{_{\mathtt{R}}}}.
\end{equation}
%%%%%%%

These are two additional conditions we have imposed so that SM Yukawa couplings are allowed for all SM fermions charged under DHC symmetry. We now solve Eqs. (\ref{Eq:anoUx1}-\ref{Eq:anoUx6}) also taking into account these two conditions. To begin with, notice that Eqs. (\ref{Eq:anoUx1}-\ref{Eq:anoUx4}) are independent of the charges of the dark fermions ($X_{f^k}$). 
As a result, solving them using Yukawa invariance conditions mentioned in Eq.\eqref{Eq:quark_singlet_charges_to_satisfy_Yukawa}, gives a unique solution in terms of SM lepton DHC charges, $X_{L}$ and $X_{e_{_{\mathtt{R}}}}$ :
%%%%%%%%%%%%
\begin{equation}\label{Eq:anomalycancel_4_eq}
X_{Q}=-\frac{X_{L}}{3},~X_{u_{_{\mathtt{R}}}}=\frac{2X_{L}}{3}-X_{e_{_{\mathtt{R}}}},~X_{d_{_{\mathtt{R}}}}= -\frac{4X_{L}}{3}+X_{e_{_{\mathtt{R}}}},~X_{_{\Phi}}=X_{L}-X_{e_{_{\mathtt{R}}}}.
\end{equation}
%%%%%%%%%%%%
Hence, we have two free parameters $X_{L}$ and $X_{e_{_{\mathtt{R}}}}$, along with $m$ DFs. The relation between them is given by Eq.\eqref{Eq:anoUx5} and Eq.\eqref{Eq:anoUx6}. Solving these two equations, using Eq.\eqref{Eq:anomalycancel_4_eq}, gives
%%%%%%%%%%%%%%%%
\begin{subequations}\label{lowcondition}
\begin{align}
\sum_{k=1}^{m} (X_{f^{^{k}}})^3=& ~ n (2X_{L} - X_{e_{_{\mathtt{R}}}})^{3},\\
\sum_{k=1}^{m} X_{f^{^{k}}}=& ~ n (2X_{L} - X_{e_{_{\mathtt{R}}}})\,,
\end{align}
\end{subequations}
%%%%%%%%%%%%%%%
where $n=1$ for SI,  $n=2$ for SII and $n=3$ for SIII.
In this work, we discuss the solutions of Eq. \eqref{Eq:anomalycancel_4_eq} and Eq. \eqref{lowcondition}, considering three DFs, i.e., we take $m=3$. Other possible solutions will be considered in follow-up works. 
The possible solutions (for $m = 3$) for the cases SI, SII, and SIII are given in Table \ref{tab:AnoConSI}, Table \ref{tab:AnoConS2}, and Table \ref{tab:AnoConS3} respectively.

In Table \ref{tab:AnoConSI}, we display the solution for the first scenario SI ($n=1$). Here, $\kappa, X_{L}$, and $X_{e_{_{\mathtt{R}}}}$ are free parameters. In this case, charges of two of the DFs mutually cancel anomalies induced by each other, and the third DF charge is fixed by the anomaly cancellation condition given in Eq. \eqref{lowcondition}.

%%%%%%%%%%%%%%%%%%%%%%%%%%%%%%%%%%%%%%%%%%%%%%%%%%%%%%%%%%%%%%%%%%%%%%%%%%%%%%%%%%%%%
\begin{table*}[ht]
\begin{center}
\begin{adjustbox}{max width=\textwidth}
\renewcommand{\arraystretch}{1.5}
 \begin{tabular}{|@{\hspace{8pt}} c  @{\hspace{6.5pt}}|@{\hspace{6.5pt}} c @{\hspace{6.5pt}}|@{\hspace{6.5pt}} c@{\hspace{6.5pt}}|@{\hspace{6.5pt}} c@{\hspace{6.5pt}}| @{\hspace{6.5pt}} c @{\hspace{6.5pt}}|@{\hspace{6.5pt}} c@{\hspace{6.5pt}}| @{\hspace{6.5pt}} c @{\hspace{6.5pt}}|@{\hspace{6.5pt}}c @{\hspace{6.5pt}}|@{\hspace{6.5pt}}c@{\hspace{8pt}}|}
 \hline 
 $Q$ & $u_{_{\mathtt{R}}}$ & $d_{_{\mathtt{R}}}$ & $L$ & $e_{_{\mathtt{R}}}$ & $f^{1}$ & $f^{2}$ & $f^{3}$& $\Phi$  \\ 
 \hline
  \hline  
 $\frac{-X_{L}}{3}$ & $\frac{2X_{L}}{3}-X_{e_{_{\mathtt{R}}}}$ & $\frac{-4X_{L}}{3}+X_{e_{_{\mathtt{R}}}}$ & $X_{L}$ & $X_{e_{_{\mathtt{R}}}}$ & $\kappa$ & $-\kappa$ & $2X_{L}-X_{e_{_{\mathtt{R}}}}$ &$X_{L}-X_{e_{_{\mathtt{R}}}}$\\
\hline
 \end{tabular}
 \end{adjustbox}  
 \end{center}
\caption{ SI $(n=1)$ case: charges of particles under DHC symmetry with three DFs $(m=3)$. See text for more details.}
\label{tab:AnoConSI}
\end{table*}
%\FloatBarrier
%%%%%%%%%%%%%%%%%%%%%%%%%%%%%%%%%%%%%%%%%%%%%%%%%%%%%%%%%%%%%%%%%%%%%%%%%%%%%%%%%%%%%

In Table \ref{tab:AnoConS2}, we presented two distinct solutions of  Eq. \eqref{Eq:anomalycancel_4_eq} and Eq. \eqref{lowcondition} with $m=3$ and $n=2$ i.e. for the second scenario (SII). In the first solution, anomalies introduced by the DFs undergo mutual cancellation, and the second solution for charges of SM fermions is analogous to the solution of SI.

%%%%%%%%%%%%%%%%%%%%%%%%%%%%%%%%%%%%%%%%%%%%%%%%%%%%%%%%%%%%%%%%%%%%%%%%%%%%
%%%%%%%%%%%%%%%%%%%%%%%%%%%%%%%%%%%%%%%%%%%%%%%%%%%%%%%%%%%%%%%%%%%%%%%%%%%%%%%%%%%%%
\begin{table*}[ht]
\begin{center}
\begin{adjustbox}{max width=\textwidth}
\renewcommand{\arraystretch}{1.5}
 \begin{tabular}{|@{\hspace{8pt}} c  @{\hspace{6.5pt}}|@{\hspace{6.5pt}} c @{\hspace{6.5pt}}|@{\hspace{6.5pt}} c@{\hspace{6.5pt}}|@{\hspace{6.5pt}} c@{\hspace{6.5pt}}| @{\hspace{6.5pt}} c @{\hspace{6.5pt}}|@{\hspace{6.5pt}} c@{\hspace{6.5pt}}| @{\hspace{6.5pt}} c @{\hspace{6.5pt}}|@{\hspace{6.5pt}}c @{\hspace{6.5pt}}|@{\hspace{6.5pt}}c@{\hspace{8pt}}|}
 \hline 
 $Q$ & $u_{_{\mathtt{R}}}$ & $d_{_{\mathtt{R}}}$ & $L$ & $e_{_{\mathtt{R}}}$ & $f^{1}$ & $f^{2}$ & $f^{3}$& $\Phi$  \\ 
 \hline
  \hline  
$\frac{-X_{L}}{3}$ & $\frac{-4X_{L}}{3}$ & $\frac{2X_{L}}{3}$ & $X_{L}$ & $2X_{L}$ & $0$ &$\kappa$& $-\kappa$ & $-X_{L}$  \\
\hline
$\frac{-X_{L}}{3}$ & $\frac{2X_{L}}{3}-X_{e_{_{\mathtt{R}}}}$ & $\frac{-4X_{L}}{3}+X_{e_{_{\mathtt{R}}}}$ & $X_{L}$ & $X_{e_{_{\mathtt{R}}}}$ & $0      $ & $2X_{L}-X_{e_{_{\mathtt{R}}}}$ & $2X_{L}-X_{e_{_{\mathtt{R}}}}$ &$X_{L}-X_{e_{_{\mathtt{R}}}}$\\
\hline
 \end{tabular}
 \end{adjustbox}  
 \end{center}
\caption{SII $(n=2)$ case: charges of particles under DHC symmetry with three DFs $(m=3)$. Each row represents the charges of the particles for a distinct solution.
See text for more details.}
\label{tab:AnoConS2}
\end{table*}
%%%%%%%%%%%%%%%%%%%%%%%%%%%%%%%%%%%%%%%%%%%%%%%%%%%%%%%%%%%%%%%%%%%%%%%%%%%%%%%%%%%%%
%%%%%%%%%%%%%%%%%%%%%%%%%%%%%%%%%%%%%%%%%%%%%%%%%%%%%%%%%%%%%%%

Table \eqref{tab:AnoConS3} displays the possible solutions of Eq. \eqref{Eq:anomalycancel_4_eq} and Eq. \eqref{lowcondition}  for the third scenario (SIII), which is the flavor universal case with $n=3$ and three DFs ($m=3$). Each row corresponds to a different solution, where $X_{L}$, $\kappa$, and $s$ are free parameters.
%%%%%%%%%%%%%%%%%%%%%%%%%%%%%%%%%%%%%%%%%%%
\begin{table*}[ht]
\begin{center}
\begin{adjustbox}{max width=\textwidth}
\renewcommand{\arraystretch}{3}
 \begin{tabular}{|@{\hspace{2.5pt}}  c  @{\hspace{2.5pt}}|@{\hspace{2.5pt}} c @{\hspace{2.5pt}}|@{\hspace{2.5pt}} c@{\hspace{2.5pt}}|@{\hspace{2.5pt}} c@{\hspace{2.5pt}}| @{\hspace{2.5pt}} c @{\hspace{2.5pt}}|@{\hspace{2.5pt}} c@{\hspace{2.5pt}}| @{\hspace{2.5pt}} c @{\hspace{2.5pt}}|@{\hspace{2.5pt}}c @{\hspace{2.5pt}}|@{\hspace{2.5pt}}c@{\hspace{2.5pt}}|}
 \hline 
 $Q$ & $u_{_{\mathtt{R}}}$ & $d_{_{\mathtt{R}}}$ & $L$ & $e_{_{\mathtt{R}}}$ & $f^{1}$ & $f^{2}$ & $f^{3}$& $\Phi$  \\ 
 \hline
  \hline
  $\frac{-X_{L}}{3}$ & $\frac{-4X_{L}}{3}$ & $\frac{2X_{L}}{3}$ & $X_{L}$ & $2X_{L}$ & $0$ &$\kappa$& $-\kappa$ &$-X_{L}$\\
  \hline
 $-\frac{X_{L}}{3}$ & $-\frac{4X_{L}}{3}+\kappa$ & $\frac{2X_{L}}{3}-\kappa$ & $X_{L}$ & $2X_{L}-\kappa$ & $\kappa$ &$\kappa$& $\kappa$ &$\kappa - X_{L}$ \\
 \hline
 $\frac{1}{s}$ &$-(\kappa-\frac{4}{s})$  &  $\kappa-\frac{2}{s}$ & $-\frac{3}{s}$ & $\kappa-\frac{6}{s}$ & $5\kappa$ & $-4\kappa$ & $-4\kappa$ & $-(\kappa-\frac{3}{s})$ \\
 \hline

 $-\frac{X_{L}}{3}$ &
 
$\begin{aligned}
 \frac{-4X_{L}}{3}\\
-\frac{s^{2}-\kappa^{2}}{8}
 \end{aligned}$  
 
 & 
$ \begin{aligned}
 \frac{2X_{L}}{3}+\\
 \frac{s^{2}-\kappa^{2}}{8}\\
 \end{aligned}$
 
 &$X_{L}$ & 

$ \begin{aligned}
 2X_{L}+\\
 \frac{s^{2}-\kappa^{2}}{8}\\
 \end{aligned}$

  &
 
 $\frac{1}{8}(5s^{2}+3\kappa^{2})$ & 

  $ \begin{aligned}
 \frac{-(4s^{2}+ 3s\kappa)}{8}\\
  \frac{-\kappa^{3}}{8s}\\
 \end{aligned}$

 & 
$ \begin{aligned}
 \frac{(-4s^{2}+ 3s\kappa)}{8}\\
 + \frac{\kappa^{3}}{8s}\\
 \end{aligned}$

  &  $-(X_{L}+\frac{s^{2}-\kappa^{2}}{8})$  \\
\hline
 \end{tabular}
 \end{adjustbox}  
 \end{center}
\caption{SIII $(n=3)$ case: charges of particles under dark hypercharge symmetry with three DFs $(m=3)$. Each row represents the charges of the particles for a distinct solution. See text for more details.}
\label{tab:AnoConS3}
\end{table*}
%%%%%%%%%%%%%%%%%%%%%%%%%%%%%%%%%%%%%%%%%%%%%%%%%%%%%%%%%%%%%%%%%%%%%%%%%%%%%%%%%%%%%

The solution in the first row is a SM hypercharge-like solution. In this case, the new DHC symmetry can be considered analogous to the hypercharge symmetry, with the addition of DFs. To make the theory anomaly-free, it is necessary for the anomalies introduced by the DFs to undergo mutual cancellation, and hence, they have equal and opposite charges. 
In the second solution, all DFs have the same charge assignment. This solution allows to construct a Yukawa term, $\overline{L}\tilde{\Phi}f^{i}$, that preserves the DHC symmetry. Hence, in this case, the Dirac mass term of DFs is allowed by the DHC symmetry. The well-known $B-L$ vector solution, wherein the charges of the DFs are assigned as $(-1,-1,-1)$, can be obtained from this solution if we take $X_L = \kappa = -1$. Notably, by replacing the Higgs charge with $X_{\Phi}$ and defining the charges of other fermions in terms of the Higgs and DFs charges, the resulting solution embodies a linear combination of hypercharge and $B-L$, as presented in Ref. \cite{Das:2016zue}. The third row corresponds to the solutions where two generations of DFs have identical DHC charges, and one generation has different charges. Notice that if we set $s=3$ and $\kappa=1$, then it results in the chiral $B-L$ solution, where the DFs charges are $(-4, -4, 5)$ \cite{Ma:2014qra,Ma:2015mjd,Ma:2015raa,Montero:2007cd}. The last row corresponds to the solution, where all DFs have distinct charges.

Both scenarios, SI and SII, are flavor-specific scenarios where all generations of SM fermions do not carry an equal charge. These scenarios may induce tree-level flavor-changing neutral currents (FCNC) mediated by the $Z'$ boson. On the other hand, SIII represents a flavor universal scenario, where all generations of SM fermions carry equal charges under $U(1)_{X}$, and hence, this scenario does not induce any $Z'$ mediated tree-level FCNC processes.
%%%
The new neutral boson ($Z'$) associated with $U(1)_{X}$ symmetry has many interesting properties. In the rest of the work, we discuss various properties and phenomenological signatures.

%%%%%%%%%%%%%%%%%%%%%%%%%%%%%%%%%%%%%%%%%%%%%%%%%%%%%%%%%%%%%%%%%%%%%%%%%%%%%%%
\section{Dark HyperCharge Gauge Boson }\label{DHCSec}
\label{Sec:DHC_Gauge_Boson}
%%%%%%%%%%%%%%%%%%%%%%%%%%%%%%%%%%%%%%%%%%%%%%%%%%%%%%%%%%%%%%%%%%%%%%%%%%%%%%%%%

The prominent characteristic of the newly introduced DHC gauged symmetry is the presence of a novel electrically neutral gauge boson $Z'$. Let us first begin with the mass generation mechanism of this boson. 
The masses of the SM gauge bosons and $Z'$ are generated through the spontaneous symmetry breaking (SSB) of both the hypercharge and DHC symmetries. However, SM Higgs doublet $\Phi$ alone could not generate the mass of all gauge bosons, and we need at least one more scalar charged under DHC symmetry to generate $Z'$ mass. New scalars are also essential for the mass generation of SM neutrinos and BSM fermions. Here, as we are discussing a general framework, we consider $``i"$ number of new scalar fields, represented as $\chi_{i}$. These scalar fields are singlets of the SM gauge symmetries but carry a charge under the DHC symmetry. For simplicity, we also assume that there is no kinetic mixing between the SM hypercharge and DHC symmetry. Under these conditions, the most general covariant derivative is given as,  
\begin{equation}\label{cod}
D_{\mu}= \partial_{\mu} +ig_{s}T^{a}_{g}G_{\mu}^{a}+igT^{a}_{w}W^{a}_{\mu} + ig'\frac{Y}{2}B_{\mu}+i\textsl{g}_{_{x}}XC_{\mu}\,.
\end{equation}
Here,  $g$, $g'$ and $\textsl{g}_{_{x}}$ denote the gauge couplings for $SU(2)_{L}$, $U(1)_{Y}$, and DHC, respectively. The charges for $U(1)_{Y}$ and DHC are represented by $Y$ and $X$. The generators of $SU(3)_{C}$ and $SU(2)_{L}$ are given by $T^{a}_{g}$ and $T^{a}_{w}$.  All scalar fields, including the SM Higgs doublet and the $\chi_{i}$'s, obtain vacuum expectation value (vev), which leads to the breaking of the DHC symmetry. After the breaking of both the electroweak and DHC symmetry, the vev of these scalar fields can be represented as follows:
\begin{equation}\label{Vevs}
\langle \Phi \rangle = \frac{1}{\sqrt{2}}\begin{bmatrix}
0 \\
v
\end{bmatrix},~~~~ \langle \chi_{i} \rangle = \frac{v_{i}}{\sqrt{2}}~.
\end{equation} 
 After SSB the mass matrix of the gauge bosons in the basis ($B^{\mu},W_{3}^{\mu},C^{\mu}$) is given as
\begin{equation}\label{GBMmat}
\mathcal{M}^2_{_{V}}= \frac{v^{2}}{4}\begin{pmatrix}
g'^{2} & -gg' & 2g'X_{_{\Phi}}\textsl{g}_{_{x}}\\
-gg'   &  g^{2} & -2gX_{_{\Phi}}\textsl{g}_{_{x}}\\
2g'X_{_{\Phi}}\textsl{g}_{_{x}} & -2gX_{_{\Phi}}\textsl{g}_{_{x}} & 4u^{2}\textsl{g}_{_{x}}^{2}
\end{pmatrix}\,,
\end{equation}
where $u^{2}=X^{^{2}}_{_{\Phi}}+u_{\chi}^{2}/v^{2}$, and $u_{\chi}$ is defined as $u_{\chi}=\sqrt{\sum_{i}(X^{^{2}}_{\chi_{_{i}}}v_{i}^{2})}$. Here, $X_{\chi_{_{i}}}$ represents the DHC charge of the scalar fields $\chi_{i}$, and $X_{_{\Phi}}$ denotes the DHC charge of the SM Higgs doublet. 
The mass matrix in Eq. \eqref{GBMmat} can be diagonalized by an orthogonal matrix $\mathcal{O}_{(\theta,\alpha)}$, and the diagonal mass matrix are given as  $\mathcal{O}_{(\theta_{w},\alpha)} \mathcal{M}^2_{_{V}} \mathcal{O}^{^{\dagger}}_{(\theta_{w},\alpha)}$.  The mass ($A^\mu, Z^\mu, Z^{\prime \mu}$) and gauge states are related with each other as
\begin{equation}
\label{unitary matrix}
\begin{bmatrix}
A^{\mu} \\
Z^{\mu} \\
Z^{\prime\mu}
\end{bmatrix} =
\begin{bmatrix}
\cos\theta_{w} &~ \sin\theta_{w} &~0\\
-\cos\alpha \sin\theta_{w} & \cos\alpha \cos\theta_{w}
&~ \sin \alpha\\
\sin\alpha \sin\theta_{w} &~  -\sin\alpha\cos\theta_{w} &~ \cos\alpha 
\end{bmatrix}  \begin{bmatrix}
B^{\mu} \\
W_{3}^{\mu}\\
C^{\mu}
\end{bmatrix}.
\end{equation}

Following the diagonalization, one mass eigenstate is zero and is identified with photon $A^\mu$. 
The other two mass eigenstates are given as,
\begin{equation}\label{mass}
M_{Z}^{2}= \frac{v^{2}}{8}(A_{0}-\sqrt{B_{0}^{2}+C_{0}^{2}}),\,M_{Z'}^{2}=\frac{v^{2}}{8}(A_{0}+\sqrt{B_{0}^{2}+C_{0}^{2}})\,,
\end{equation}
where $A_{0}=g^{2}+{g'}^{2}+4u^{2}\textsl{g}_{_{x}}^{2},~ B_{0}=4X_{_{\Phi}}\textsl{g}_{_{x}}\sqrt{g^{2}+{g'}^{2}}$, and $C_{0}=4u^{2}\textsl{g}_{_{x}}^{2}-(g^{2}+{g'}^{2})$\,. 
The two mixing angles can be written as
\begin{equation}\label{Eq:Rotation_angle}
    \tan\theta_{w} = \frac{g'}{g},\quad \tan2\alpha = \frac{B_{0}}{C_{0}}.
\end{equation}
with $\theta_w$ same as the SM weak mixing angle. 
Note that the $W$ boson mass remains the same as in the SM and is given as $M_{W}^{2}=(gv)^{2}/4$.

The vev parameter $u_{\chi}=\sqrt{\sum_{i}(X^{^{2}}_{\chi_{_{i}}}v_{i}^{2})}$ can be rewritten in term of $Z'$ mass as
\begin{equation}\label{Eq:ux_interm_Mzp}
    u_{\chi} = \frac{M_{Z'}}{\textsl{g}_{_{x}}} \sqrt{\frac{4M_{Z'}^{2} -v^{2}(g^{2} +g'^{2} + 4\textsl{g}_{_{x}}^{2}X_{\Phi}^{2}) }{4M_{Z'}^{2} - v^{2}(g^{2}+g'^{2})}} \,.
\end{equation}
From Eqs. \eqref{mass}, \eqref{Eq:Rotation_angle}, and  \eqref{Eq:ux_interm_Mzp}, it is clear that the new gauge mixing angle, $\alpha(\textsl{g}_{_{x}},M_{Z'},X_{\Phi})$ is a function of gauge coupling $\textsl{g}_{_{x}}$, $Z'$ mass $M_{Z'}$ and Higgs DHC charge $X_{\Phi}$. 
\begin{figure*}[h]
   \centering
   \captionsetup{justification=raggedright}
  \includegraphics[width=0.7\textwidth]{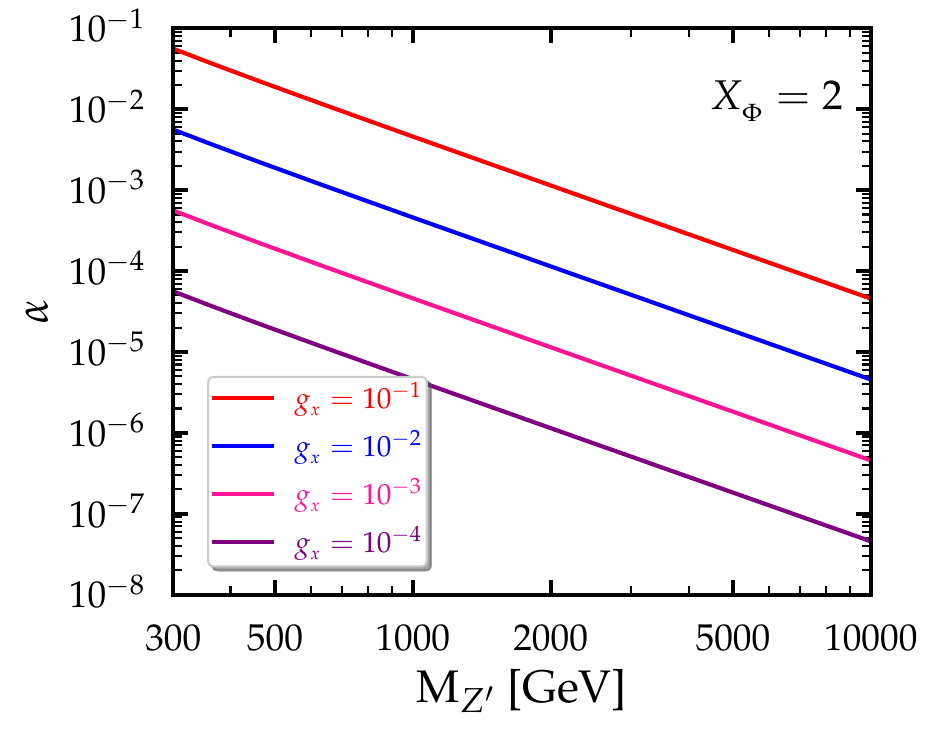}
  \caption{Gauge mixing angle $\alpha$ vs $M_{Z'}$ for varying values of $\textsl{g}_{_{x}}$. Higgs $U(1)_{X}$ charge is fixed to $X_{\Phi}=2$.  }\label{Fig:Alpha_Mzp}
\end{figure*}
In Fig. \ref{Fig:Alpha_Mzp}, we have plotted the gauge mixing angle $\alpha(\textsl{g}_{_{x}},M_{Z'},X_{\Phi})$ vs $Z'$ mass ($M_{Z'}$), with fixed $X_{\Phi}=2$ and varying values of $\textsl{g}_{_{x}}$. It is evident from Fig. \ref{Fig:Alpha_Mzp}, that as \(M_{Z'} \rightarrow \infty\), the mixing angle \(\alpha\) approaches zero. This behavior is characteristic of ``mass mixing", unlike kinetic mixing, where the mixing angle between \(Z\) and \(Z'\) can remain large even for large $M_{Z'}$ values.

\begin{figure*}[h]
   \centering
   \captionsetup{justification=raggedright}
  \includegraphics[width=0.7\textwidth]{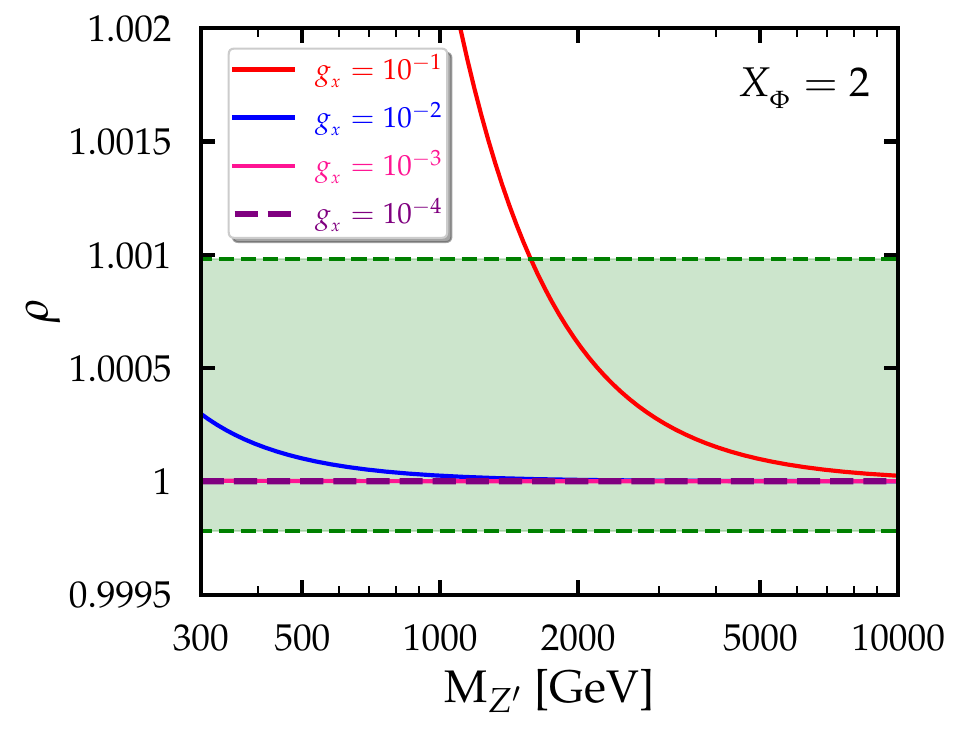}
  \caption{ $\rho$ parameter vs mass of $Z'$, $M_{Z'}$ for varying $\textsl{g}_{_{x}}$ values and fixed Higgs $U(1)_{X}$ charge $X_{\Phi} =2$.}\label{Fig:Rho_MZP_Phi2}
\end{figure*}

This $Z'$ boson has several characteristic  properties that can be probed experimentally. One such electroweak observable is the $\rho$ parameter, defined as, $\rho=M_{W}^{^{2}}/(M_{Z} {\cos}{\theta_{w}})^{2}$. This parameter is $1$ at the tree level in the SM. But in $SM \otimes U(1)_{X}$ theories  it is typically no longer $1$ at tree level.
Therefore, we can introduce a new parameter $\rho'$ that is $1$ at the tree level, defined as \cite{Bento:2023flt,Bento:2023weq},
\begin{equation}
\label{Eq:rho_prime}
    \rho' = \frac{\rho}{\cos^{2}{\alpha} + \left( \frac{M_{Z'}}{M_{Z}} \right)^{2} \sin^{2}{\alpha}}=1 \,.
\end{equation}
Hence $\rho$ parameter can be conveniently written as
\begin{equation}
\label{Eq:rho_prime}
    \rho -1 =  \left[ \left( \frac{M_{Z'}}{M_{Z}} \right)^{2}-1 \right] \sin^{2}{\alpha}.
\end{equation}
Since we have identified the heavier mass eigenstate as $Z'$ i.e. $ M_{Z'}(\textsl{g}_{_{x}},u_{\chi}) > M_{Z}(\textsl{g}_{_{x}},u_{\chi})$, therefore the mass of the $Z$ boson, $M_{Z}(\textsl{g}_{_{x}},u_{\chi})$, is found to be lower than the value predicted by the SM. Consequently, the $\rho$ parameter exceeds the value anticipated by the SM.
%
%  In this limit $\rho$ parameter could be approximated as,
% %
% \begin{equation}\label{Eq:approx_rho}
%     \rho \approx \frac{1}{1- \left(\frac{X_{\Phi}v}{u_{\chi}} \right)^{2}}\,. 
% \end{equation}
% %
% Hence, the $\rho$ parameter sets a lower bound on $u_{\chi}$, and for a given $M_{Z'}$, this lower bound on $u_{\chi}$ could be translated as the upper bound on $\textsl{g}_{_{x}}$ from Eq.\eqref{Eq:ux_interm_Mzp}.
%
In Fig. \ref{Fig:Rho_MZP_Phi2}, we 
varied $M_{Z'}$ freely and showed this constraint on $\rho-M_{Z'}$ plane for various $\textsl{g}_{_{x}}$ values. The green band corresponds to the $\rho$ parameter's current experimental value $1.00038 \pm 0.00060$ at $3\sigma$ \cite{ParticleDataGroup:2020ssz}. 
It is clear from Fig. \ref{Fig:Rho_MZP_Phi2} that the high value of gauge coupling at low $Z'$ mass is ruled out by the $\rho$ parameter. Thus, $\rho$ parameter provides important constraint on the allowed ($g_x, M_{Z'}$) values.

%%%%%%%%%%%%%%%%%%%%%%%%%%%%%%%%%%%%%%%%%%%%%%%%%%%%%%%%%%%%%%%%%%%%%%%%%%%%%%%%%%%%%
\section{$Z^{\prime}$ production and decay at colliders}\label{Sec:Zpcollider}
%%%%%%%%%%%%%%%%%%%%%%%%%%%%%%%%%%%%%%%%%%%%%%%%%%%%%%%%%%%%%%%%%%%%%%%%%%%%%%%%%%%

We now focus our attention on collider searches for the $Z'$ gauge boson. Since the $Z'$ couples with both quarks and leptons, it can be searched in high-energy colliders such as LHC through production via the Drell-Yan process and decay in the leptonic channels, i.e.
\begin{equation}
%q \,\bar{q} 
 p \, p \longrightarrow Z'\, \longleftrightarrow l^+ \, l^- \,.
\end{equation}

\begin{table}[ht]
\renewcommand{\arraystretch}{1.5}
 \begin{tabular}{|@{\hspace{5pt}} c  @{\hspace{4pt}}|@{\hspace{4pt}} c @{\hspace{4pt}}|@{\hspace{4pt}} c@{\hspace{4pt}}|@{\hspace{4pt}} c@{\hspace{4pt}}| @{\hspace{4pt}} c @{\hspace{4pt}}|@{\hspace{4pt}} c@{\hspace{4pt}}| @{\hspace{4pt}} c @{\hspace{4pt}}|@{\hspace{4pt}}c @{\hspace{4pt}}|@{\hspace{4pt}}c@{\hspace{4pt}}|@{\hspace{4pt}}c@{\hspace{4pt}}|@{\hspace{4pt}}c@{\hspace{5pt}}|}
 \hline 
$U(1)$& $Q$ & $u_{_{\mathtt{R}}}$ & $d_{_{\mathtt{R}}}$ & $L$ & $e_{_{\mathtt{R}}}$ & $f_{1}$ & $f_{2}$ & $f_{3}$  & $\Phi$ & $\chi_{0}$  \\ 
 \hline
  \hline  
$U(1)_{Y}$ & $\frac{1}{3}$ & $\frac{4}{3}$  &  $\frac{-2}{3}$ & $-1$ & $-2$ & $0$ & $0$ & $0$ & $1$ & $0$  \\
\hline
$U(1)_{X}$ & $-\frac{1}{3}$ &$\frac{5}{3}$  &  $-\frac{7}{3}$ & $1$ & $-1$ & $10$ & $-18$ & $17$ & $2$ & $-6$  \\
\hline 
 \end{tabular}  
\caption{Charges of particles under $U(1)_{Y}$ and DHC symmetry. This solution corresponds to the 4th row of Table \ref{tab:AnoConS3}  with $X_{L} = 1$, $s=1$ and $\kappa=5$. }
\label{tab:Finalcharge}
\end{table}
 All the solutions discussed in Tables. (\ref{tab:AnoConSI} - \ref{tab:AnoConS3}) lead to interesting phenomenological signatures. However, for the sake of definiteness, we need to fix the charges of particles, i.e. choose a benchmark model. For this purpose, we have selected a particular benchmark solution shown in Table \ref{tab:Finalcharge}. This solution corresponds to the 4th row of Table \ref{tab:AnoConS3} with $X_{L} = 1$, $s=1$ and $\kappa=5$.
 This is a flavor-universal solution,  where the $U(1)_{X}$ charges associated with the SM fermions bear a close resemblance to hypercharges, and the $U(1)_{X}$ charges of DFs are relatively higher than those of SM fermions. This means that the associated $Z'$ gauge boson primarily decays to DFs. 
 In the rest of this work, we will use this benchmark model for phenomenological analysis. 
 For the case of this benchmark model, the mass of the $Z'$ gauge boson can be generated by the addition of only one SM singlet scalar $\chi_{0}$ with charge $X_{\chi_0} = -6$. This particular choice is motivated by the fact that it also allows us to write the Generalized Weinberg operator $\overline{L^{c}}\Phi\Phi L \chi_{0}$~\cite{CentellesChulia:2018gwr}, which can generate effective masses for the neutrinos. The various possible Ultra Violet (UV) completions of this operator will be studied in follow-up works \footnote{Note that such UV completions can have additional scalars. This analysis remains valid as long as these scalars are heavy enough such that the decays of $Z'$ to these scalars are kinematically forbidden.}.
 
The production cross-section of the $Z'$ boson is primarily governed by the mass of $Z'$ and DHC gauge coupling strength. 
\begin{figure}[ht]
   \centering
   \captionsetup{justification=raggedright}
  \includegraphics[width=0.7\textwidth]{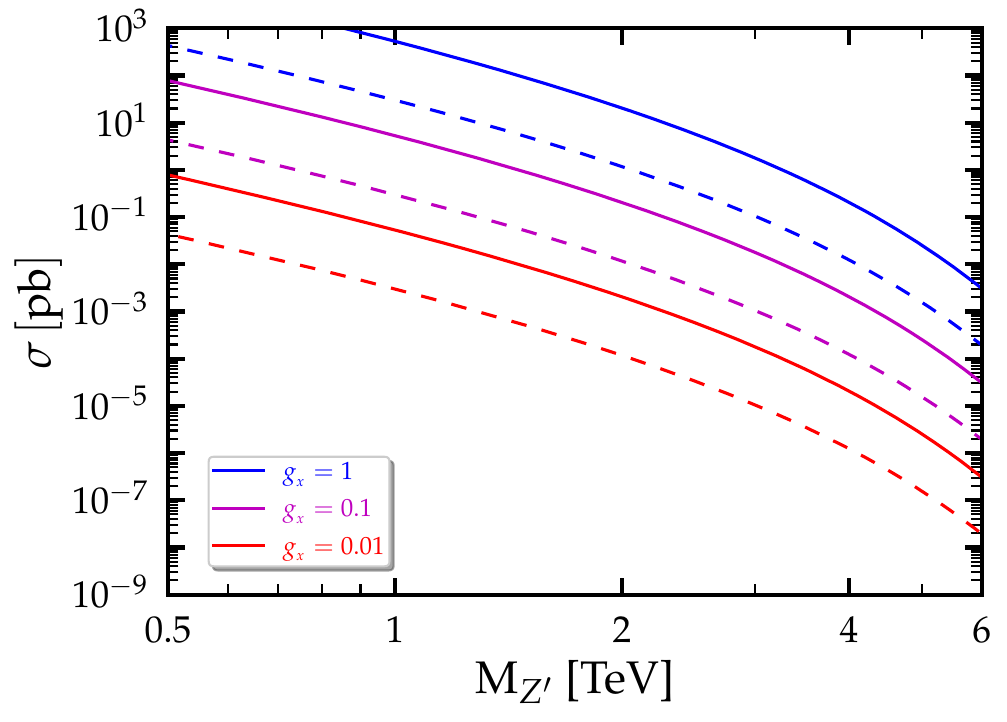}
  \caption{$Z'$ production cross-section in $pp$ collisions at $\sqrt{s}=13$ TeV for various $\textsl{g}_{_{x}}$ values. The solid lines represent DHC symmetry, while dashed lines represents $B-L$ symmetry.}\label{producCrc}
\end{figure}
In Fig. \ref{producCrc}, we illustrate the $Z'$ production cross-section in $pp$ collisions at $\sqrt{s}=13$ TeV for various $\textsl{g}_{_{x}}$ values. The solid lines denote the production cross-section for our benchmark DHC model. For comparison, we have also plotted the analogous production cross-section for the vector $B-L$ symmetry, shown by dashed lines. It is important to note that the quark charges in our benchmark DHC model are higher than those for the $B-L$ symmetry. Thus, for any given value of $\textsl{g}_{_{x}}$, the production cross-section for DHC $Z'$ is always higher than that of $B-L$, as can be clearly observed in the plot. 

Having analyzed the production of the $Z'$ at a hadron collider, we now look at its decay modes. The decay modes and branching ratios of the $Z'$ boson are determined by its coupling with other particles. The analytical expressions for partial decay widths of the $Z'$ boson into the relevant final states are shown in Appendix \ref{APn3}. 
For our benchmark DHC model, the DFs have relatively higher charges compared to the SM fermions. 
This results in a larger branching fraction of the $Z'$ boson decaying into DFs. Since the DFs belong to the dark sector, this leads to a large
invisible decay of the $Z'$ boson.

We categorize $Z'$ two body decays as either visible or invisible. Visible decays are those that leave a detectable signature on the detectors, such as charged leptons, heavy quarks, and jets. 
\begin{figure}[ht]
\begin{center}
\includegraphics[width=0.7\textwidth]{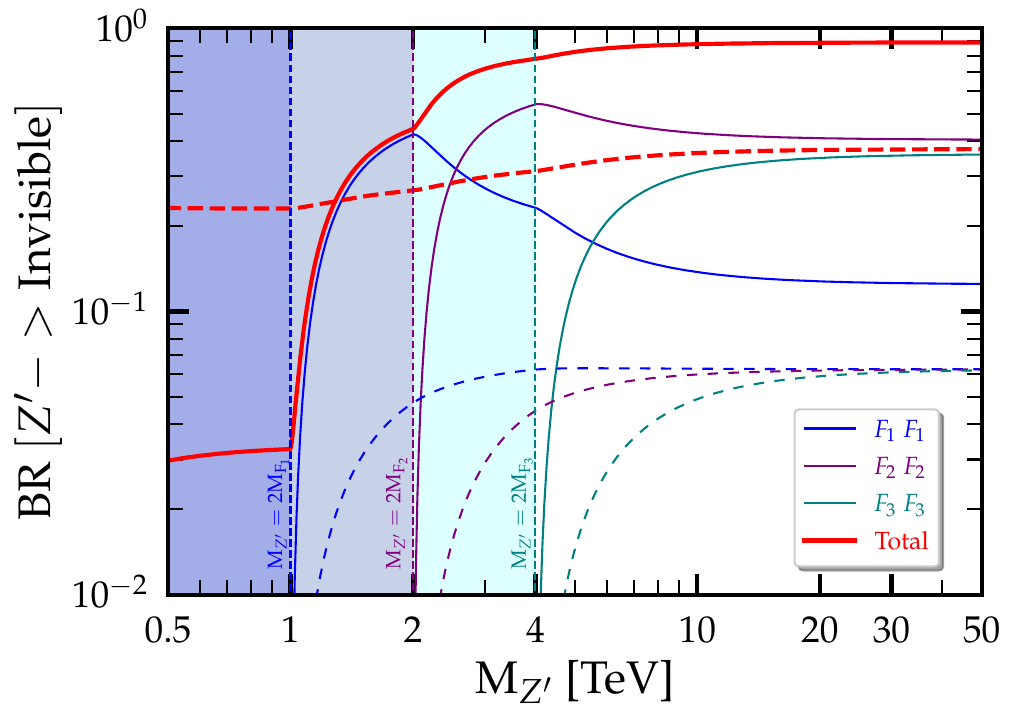}
\end{center}
\caption{$Z'$ invisible branching fraction versus $M_{Z'}$. The solid (dashed) lines represent DHC ($B-L$) symmetries. DF masses are set at 0.5 TeV, 1 TeV, and 2 TeV. Compared to $B-L$ case, the DHC $Z'$ has much larger invisible branching fraction.}
\label{BrInvisible}
\end{figure}
In contrast, invisible decays are those that exit the detector without being detected.  This is typically because they are electrically neutral and do not have any visible particle in the final state, against whose momentum some amount of missing transverse momentum ($P_{T}$ ) can be reconstructed. This includes the two body decays of $Z'$ into SM neutrinos and DFs.
Figure \ref{BrInvisible} illustrates the two-body invisible decay channels of $Z'$. Here, $F_{i}$ represents the mass eigenstates of DFs. The solid lines correspond to the benchmark model for DHC symmetry, while the dashed lines depict $B-L$ symmetry, which are plotted for the sake of comparison. In both the cases we have shown the $Z'$ branching fractions for fixed DF masses of $500$ GeV, $1000$ GeV, and $2000$ GeV. The vertical dashed lines correspond to $M_{Z'}= 2M_{F_{i}}$, beyond which the decay channels into that specific DF open up. The red line represents the total invisible decay, encompassing both DFs and SM neutrinos. 
The extreme left of the plot represents scenarios where $Z'$ does not decay into DFs, and only SM neutrinos contribute to branching fraction. The extreme right indicates regions where the $M_{F_{i}}$ masses are negligible compared to $M_{Z'}$, causing the branching fraction to saturate. Due to the high $U(1)_{X}$ charges of DFs, the invisible branching fraction of $Z'$ is notably high in DHC symmetry, reaching approximately $90\%$ at saturation. Conversely, under $B-L$ symmetry, it is around $38\%$.

\begin{figure}[ht]
\centering
\captionsetup{justification=raggedright}
\includegraphics[width=0.7\textwidth]{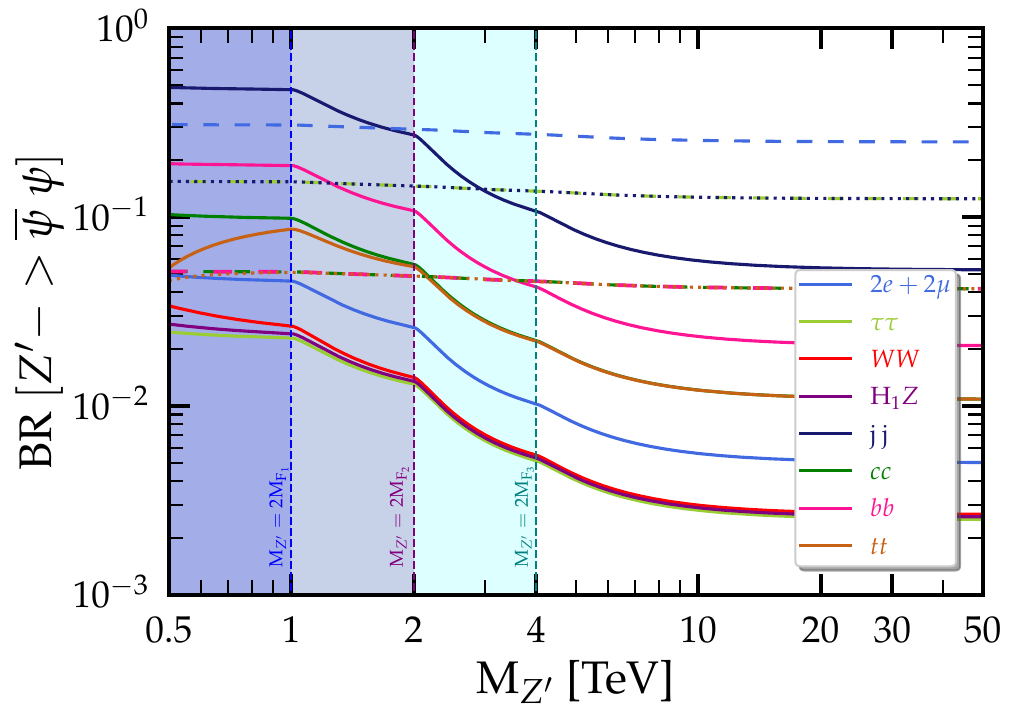}
\caption{$Z'$ branching fraction to visible modes vs $Z'$ mass. The solid lines represent DHC symmetry, while the dashed represent $B-L$ symmetry.}
\label{VisInvis}
\end{figure}
Figure \ref{VisInvis}, illustrates the visible channels of $Z'$ two-body decays. As before, the solid lines correspond to the benchmark model for DHC symmetry, while the dashed represents $B-L$ symmetry.   As evident from Fig. \ref{VisInvis}, when $Z'$ decay to all DFs is kinematically allowed, the branching fractions of $Z'$ into visible channels are significantly less compared to the $B-L$ model. Specifically, in the fermionic decay modes, the dileptonic branching fraction at saturation, which is maximum in $B-L$ symmetry ($25 \%$), is minimum in DHC symmetry ($ 0.5 \%$). In fact, in the DHC case the branching fraction to visible channels is merely $10\%$ compared to $ 62 \%$ for the $B-L$ case.
\begin{figure}[ht]
\begin{center}
\includegraphics[width=0.7\textwidth]{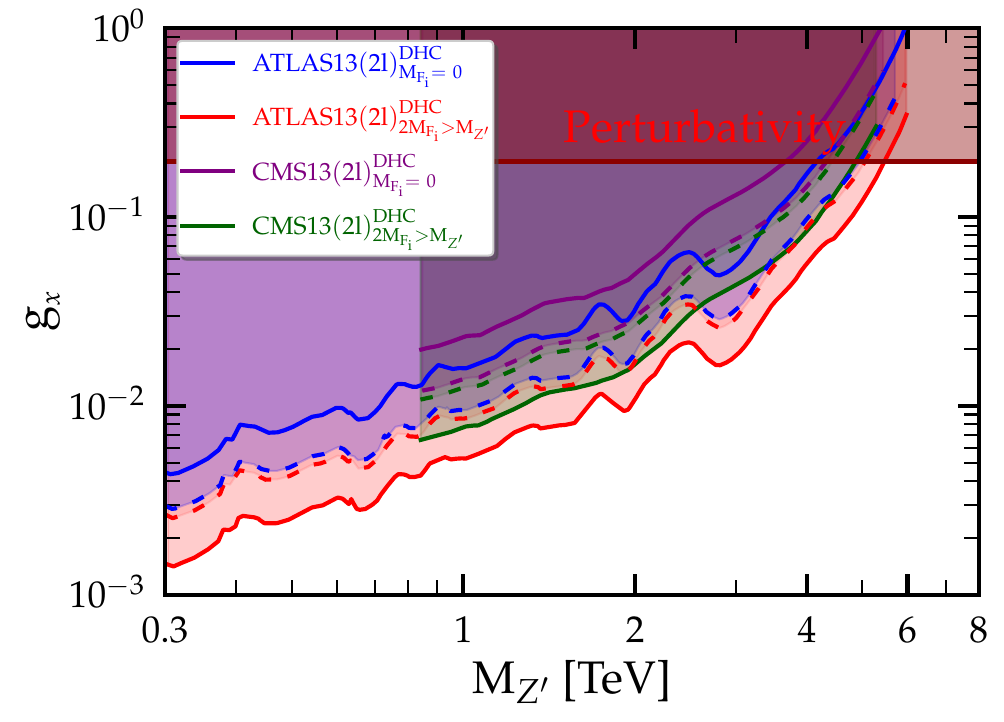}
\end{center}
\caption{The constraint on ($g_{x}-M_{Z'}$) plane. Shaded areas indicate regions excluded by ATLAS and CMS dilepton searches. The solid (dashed) lines represent DHC ($B-L$) symmetries.}
\label{gxconstaints}
\end{figure}

Using the production and decay information of the $Z'$, we can derive the constraints on the $\textsl{g}_{_{x}} - M_{Z'}$ parameter space for our benchmark DHC model as shown in Fig.~\ref{gxconstaints}.
In plotting Fig. \ref{gxconstaints} we have used the ATLAS search for $Z'$ in dilepton resonance at $pp$ collisions with $\sqrt{s}=$ 13 TeV and an integrated luminosity of 139 fb$^{-1}$ \cite{ATLAS:2019erb}. Additionally, we incorporated results from CMS, from a data set of proton-proton collisions at $\sqrt{s}=13$ TeV from 2016 to 2018, corresponding to a total integrated luminosity of up to $140~ \text{fb}^{-1}$ \cite{CMS:2021ctt}.

We have presented these limits under two scenarios: one where the decay of $M_{Z'}$ to DFs is kinematically forbidden ($M_{F_i} > M_{Z'}/2$), and another where $M_{F_i} << M_{Z'}/2$. 
In Fig.~\ref{gxconstaints}, the limits obtained from ATLAS dilepton search at $\sqrt{s}=13$ TeV (ATLAS13($2l$)) are shown in red and blue solid lines corresponding to the first and second scenarios, respectively. The solid green and purple lines depict the analogous constraints obtained using CMS data.
For comparison, we have also plotted the limits for vector $B-L$ symmetry, represented by the corresponding dashed lines. 
%N
Note that compared to the $B-L$ case, in our benchmark DHC model, the $Z'$ has higher production rate but also larger invisible branching fraction.
This leads to constraints on  $\textsl{g}_{_{x}} - M_{Z'}$ which can be either more constrained ($M_{F_i} > M_{Z'}/2$) or more relaxed ($M_{F_i} << M_{Z'}/2$) than the $B-L$ case as shown in Fig.~\ref{gxconstaints}. Finally, the horizontal red line is the constraint obtained from tree-level perturbativity by the demand that $X_{f^2} \textsl{g}_{_{x}} \leq \sqrt{4\pi}$, where $X_{f^2}$ is the largest $U(1)_X$ charge in our benchmark DHC model, see Tab.~\ref{tab:Finalcharge}. 
Additional constraints on a chiral $Z'$ arise from atomic parity violation \cite{Mahanthappa:1991pw,Diener:2011jt} and forward–backward asymmetry (FBA) \cite{Accomando:2015cfa}. However, the bounds derived from these observables are typically weaker than those obtained from the resonance searches discussed here.

In summary, the $Z'$ in our benchmark DHC model couples strongly with quarks and thus can be readily produced at hadron colliders. It also has very characteristic decay pattern with large invisible branching fractions. This allows one to also distinguish it from other $Z'$ at collider experiments.

\FloatBarrier
%%%%%%%%%%%%%%%%%%%%%%%%%%%%%%%%%%%%%%%%%%%%%%%%%%%%%%%%%%%%%%%%%%%%%%%%%%%%%%%%%%%%%
\section{The Dark Sector}
\label{Sec:DM}
%%%%%%%%%%%%%%%%%%%%%%%%%%%%%%%%%%%%%%%%%%%%%%%%%%%%%%%%%%%%%%%%%%%%%%%%%%%%%%%%%%%%%

As we mentioned earlier, the new fermions $f^i$ added for anomaly cancellation belong to the dark sector with the lightest of them a potential DM candidate. In this section, we look at the dark sector in detail and show that the lightest DF can satisfy all the DM constraints.  We consider the DFs to be Majorana fermions, and their mass term, in general, can be obtained from Yukawa terms\footnote{Note that depending on the DHC charges of the DFs, for some benchmark models, the DFs can even have gauge invariant mass terms of the form $\frac{M_{ij}}{2} \overline{f_{i}^{c}}f_{j}$.},
\begin{equation} \label{Eq:DFmass}
-\frac{Y_{ij}}{2} \overline{f_{i}^{c}}f_{j} \chi_r + \text{h.c.}\,,
\end{equation}
\begin{figure*}[bh]
\begin{center}
\includegraphics[width=0.65\textwidth]{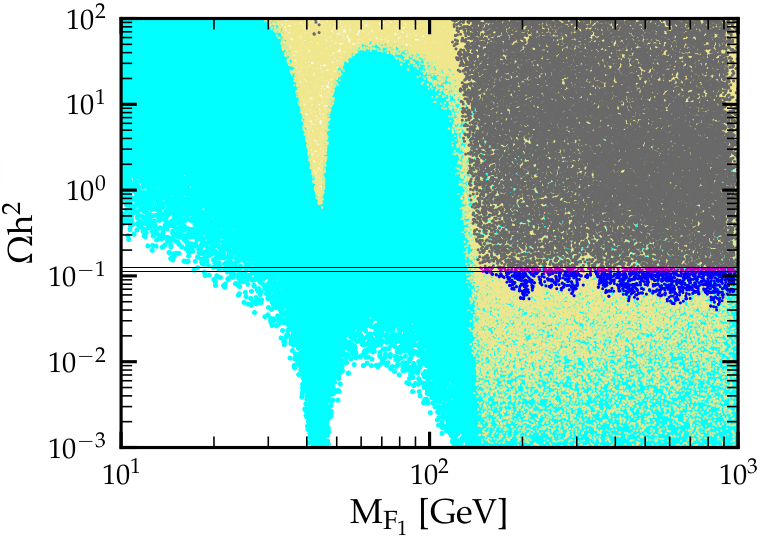}
\end{center}
\caption{ DM relic density plotted against DM mass. The cyan points are ruled out by $\rho$ parameter constraints whereas the yellow points are ruled out by the collider constraints. The gray (blue) points satisfy both $\rho$ parameter and  collider constraints but lead to DM over (under) abundance. Magenta points satisfy all the constraints including the correct DM relic abundance.
For more details see the main text.}
\label{RelicP}
\end{figure*}
 where $i,j=1,2,3$, and $\chi_r$; $r = 1,2, \cdots$ are SM gauge  singlet scalars carrying appropriate DHC charges. After SSB, the $\chi_r$ vevs will lead to mass generation for the DFs with the DF mass matrix given by $ M_{ij} = Y_{ij} \langle \chi_r \rangle $.  After diagonalization, we get the physical DFs, $F_{i}$: $i = 1,2,3$. For the sake of definiteness, here we consider DFs to be nearly degenerate with the hierarchy $M_{F_{1}}<M_{F_{2}}<M_{F_{3}}$. The stable DM candidate is the lightest DF, i.e. $F_1$. To write the DFs mass term as shown in Eq. \eqref{Eq:DFmass}, the DHC charges of the scalars can be assigned in various ways. One option is to introduce three scalars, each with charges that are double the DHC charges of each corresponding DF. Alternatively, cross terms can be introduced between different DFs by assigning scalars a DHC charge equal to the sum of the DHC charges of the two different DFs. Clearly, depending on a given model there are several combinations of scalars with appropriate DHC charges that can be used. 

 For the benchmark model under discussion, for further analysis, we take one such choice and add three scalars with DHC charges $(-20, 8, 1)$.
To avoid a massless Goldstone mode, we further add a scalar with $U(1)_X$ charge 2.
Note that other choices of scalars with appropriate DHC charges are equally viable. 
 Since, the scalar sector has several options, in order to be as model-independent as possible, we 
 assume that the additional scalars are quite heavy and/or couple very weakly to the $SU(2)_L$ doublet scalar $\Phi$. 
 This ensures that the DM phenomenology is dominated by the DF's gauge coupling to $Z'$. This makes our analysis in this section somewhat model-independent as the $Z'$ interaction to DM is dictated solely by the DHC symmetry.
 It has the added advantage that the lightest 
 CP-even scalar, $H_{1}$, can be identified as the 125 GeV scalar. It will be primarily composed of the neutral component of $\Phi$ with a very small admixture from other scalars and can thus trivially satisfy the LHC bounds on its couplings.
Hence, the only free parameters relevant for DM phenomenology are $\textsl{g}_{_{x}}$ and $M_{Z'}$. We varied them within the ranges $\textsl{g}_{_{x}} = [ 10^{-4} - 1 ]$ and $M_{Z'} = [ 300 - 1000 ]$ GeV. All the relevant Feynman diagrams are shown in the Appendix \ref{APn1}.

In Fig. \ref{RelicP}, we show the relic density vs DM mass. 
\begin{figure}[h!]
\begin{center}
\includegraphics[width=0.7\linewidth]{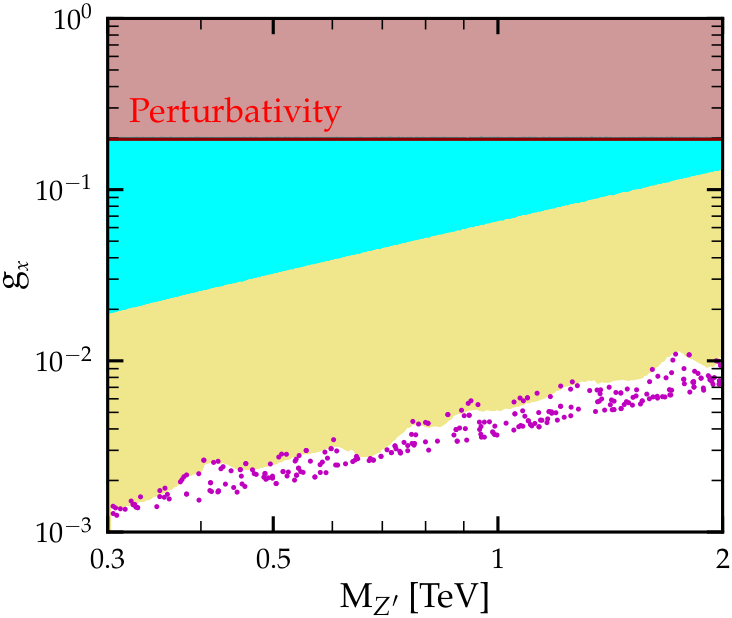}
\end{center}
\caption{ $Z'$ mass vs its gauge coupling $\textsl{g}_{_{x}}$. The red-shaded region is excluded by perturbativity constraints. The cyan (yellow) points are ruled out by the $\rho $ parameter (collider constraints). Magenta points are within 3 $\sigma$ of Planck's cold DM range and also satisfy the DM-nucleons elastic scattering constraints.  }
\label{fig:ConsRelic}
\end{figure}
The prominent dip occurring around 45.5 GeV, corresponds to $M_{F_{1}} \approx M_{Z}/2$. Cyan points are ruled out by the $\rho$ parameter limits. 
From Fig.~\ref{fig:ConsRelic}, it can be seen that the $\rho$ parameter imposes an upper bound on $\textsl{g}_{_{x}}$ for a given $M_{Z'}$.
Consequently, the $\rho$ parameter rules out many low relic density points with high $\textsl{g}_{_{x}}$. The yellow points are ruled out by collider constraints discussed in Sec.~\ref{Sec:Zpcollider}. The blue (gray) points satisfy the above two constraints but lead to under-abundant (over-abundant) DM relic density. The magenta points satisfy the $3\sigma$ allowed range for the cold DM derived from the Planck satellite data, $0.1126 \leq \Omega h^{2} \leq 0.1246$ ~\cite{Planck:2018vyg} along with satisfying all the other constraints.  We found that in the low mass range of DM ($ \lesssim 150~\text{GeV}$), the constraints on $\textsl{g}_{_{x}}$ imposed by the $\rho$ parameter and collider data do not permit the fulfillment of relic density conditions, even at the $Z$ resonance. Our DM $F_{1}$ satisfies  the relic density constraints at $M_{F_{1}} \gtrsim 150$ GeV. It should be noted that this limit on DM mass is obtained when the DM annihilation happens predominantly through the gauge bosons. If the scalars are taken light, then new DM decay channels will open up, potentially lowering the lower limit on DM mass.

\begin{figure}[h]
\begin{center}
\includegraphics[width=0.7\textwidth]{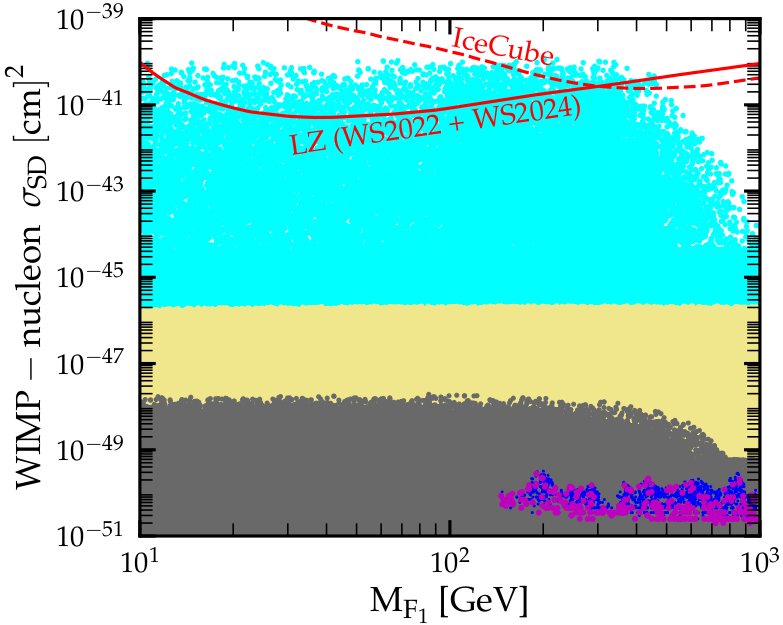}
\end{center}
\caption{Spin-dependent WIMP-nucleon cross section versus DM mass, following the color scheme of Fig. \ref{RelicP}. Points above the dashed red line are excluded by IceCube \cite{IceCube:2016dgk}, and those above the solid red line are excluded by LZ direct detection experiment~\cite{LZCollaboration:2024lux}.}
\label{DD}
\end{figure}

Finally, we look at the constraints from dark matter direct detection experiments \cite{LZCollaboration:2024lux,PandaX:2024qfu, XENONCollaboration:2023orw}  as well as constraints from DM search by neutrino experiments  \cite{IceCube:2016dgk,ANTARES:2016xuh,Frankiewicz:2015zma}.
In our case the Majorana DM, $F_{1}$ can interact with nucleons through tree-level $Z$ and $Z'$ exchanges. The relevant Feynman diagrams are provided in Appendix \ref{APn1}. These interactions are spin-dependent, as vector coupling of Majorana fermions with gauge bosons is identically zero. 
We found that, in our case, the
IceCube \cite{IceCube:2016dgk} and the LUX-ZEPLIN (LZ) \cite{LZCollaboration:2024lux} collaborations impose the most stringent constraints. 
In Fig.\ref{DD}, we show the DM-nucleon elastic scattering cross-section vs DM mass. 

The colour code has the same meaning as Fig.\ref{RelicP}.
The most recent upper bound from the LZ and IceCube collaborations is shown by
the solid and dashed red lines, respectively. We found that all the magenta points that satisfy the relic density and other constraints also satisfy the constraints from LZ and IceCube. Thus, the BSM fermions needed for anomaly cancellations can indeed be considered as belonging to the dark sector, with the lightest of them satisfying all DM constraints.

\FloatBarrier

%%%%%%%%%%%%%%%%%%%%%%%%%%%%%%%%%%%%%%%%%%%%%%%%%%%%%%%%%%%%%%%%%%%%%%%%%%%%%%%%%%%%%%%%

\section{conclusion} \label{Sec:Conclusion}
%%%%%%%%%%%%%%%%%%%%%%%%%%%%%%%%%%%%%%%%%%%%%%%%%%%%%%%%%%%%%%%%%%%%%%%%%%%%%%% 

We have examined chiral solutions in which SM fermions are charged non trivially under the $U(1)_{X}$ symmetry. Gauge anomaly cancellation is achieved by introducing new SM gauge singlet right-handed dark fermions charged under $U(1)_{X}$ symmetry. We systematically present chiral solutions under three different scenarios :
\begin{itemize}
     \item  Only one generation of SM fermions are charged under $U(1)_{X}$ symmetry.
    \item  Two generations of SM fermions of each type share the same charge under $U(1)_{X}$ symmetry, while one generation remains uncharged.
     \item  All three generations of SM fermions are charged under the new symmetry, and charges of a given fermion type are identical across generations.
\end{itemize}   
We refer to this class of chiral symmetries as ``the dark hyperCharge'' symmetries because the new BSM fermions introduced for anomaly cancellation belong to the dark sector, with the lightest among them being a good dark matter candidate. Furthermore, the $Z'$ gauge boson associated with such $U(1)_X$ symmetries mediates interactions between the SM and the dark sector.  

As both leptons and quarks are charged under the $U(1)_{X}$ symmetry, the $Z'$ boson could be searched in high-energy collider experiments. To study the viable parameter space, we first select a benchmark model and investigate the collider constraints on the $Z'$ boson in the heavy $Z'$ scenario ($M_{Z'} > M_{Z}$). In this benchmark model, the dark fermions possess higher $U(1)_{X}$ charges than the SM fermions, leading to a significant invisible branching fraction for the $Z'$. Using the production and decays of $Z'$ boson into leptonic channels, we presented the constraints on the $Z'$ mass and gauge coupling derived from hadronic collider data.
Subsequently, we explored the potential of the lightest dark fermion  to be the dark matter  candidate.  Our analysis indicates that the lightest dark fermion, denoted as $F_{1}$, can indeed serve as a DM candidate satisfying all relevant DM properties as well as the current experimental constraints.
The complementary case $M_{Z'} < M_Z$ is explored in our follow-up work~\cite{Majumdar:2024dms}. 

Our results have important implications for both model building and phenomenology. In addition to highlighting a largely unexplored class of $U(1)_X$ extensions, our work provides a novel connection between the visible and dark sectors. Unlike well-known vector $U(1)_X$ models, where there are only a few options for the charges of SM fermions, our chiral models allow for a much broader range of possibilities. As we discussed, the ratios of $U(1)_X$ charges among quarks, leptons, and dark fermions vary across different realizations. This flexibility makes these models well-suited for potentially addressing various unexplained experimental anomalies, such the ATOMKI~\cite{Krasznahorkay:2015iga,Krasznahorkay:2021joi,Krasznahorkay:2022pxs,Barducci:2022lqd}, PADME~\cite{Bossi:2025ptv,Arias-Aragon:2025wdt}, KTeV~\cite{KTeV:2006pwx}, the electron $g-2$ ~\cite{Parker:2018vye}, and $B \to K \nu \nu$ anomaly~\cite{Belle-II:2023esi}.

%%%%%%%%%%%%%%%%%%%%%%%%%%%%%%%%%%%%%%%%%%%%%%%%%%%%%%%%%%%%%%%%%%%%%%%%%%%%%%%%%%%%%%%%
\begin{acknowledgments}
The work of H.P. is supported by the Prime Minister Research Fellowship (ID: 0401969). The work of R.S.  is supported by the Government of India, SERB Startup Grant SRG/2020/002303.  We use the package SARAH-4.14.5 \cite{Staub:2015kfa} to calculate vertices and mass matrices. SPheno-4.02 \cite{Goodsell:2017pdq,Porod:2003um} handled numerical calculations, and micrOMEGAS-5.2.13 \cite{Belanger:2020gnr,Belanger:2014vza} is used for computing DM properties. We also used MadGraph for collider simulations \cite{Alwall:2014hca}. \\

RS dedicates this work to his wife and their ten-month-old daughter, Akanksha, who endured many long hours of his absence during its completion—a process that took more than three times the duration of Akanksha’s gestation.

\end{acknowledgments}

\vspace{2cm}

\appendix

%%%%%%%%%%%%%%%%%%%%%%%%%%%%%%%%%%%%%%%%%%%%%%%%%%%%%%%%%%%%%%%%%%%%%%%%%%%%%%%%%%%%%%%%%%
\section{Partial decay widths of the  $Z'$ boson}
\label{APn3}
%%%%%%%%%%%%%%%%%%%%%%%%%%%%%%%%%%%%%%%%

The various two body decay modes of the $Z'$ gauge boson associated with a given $U(1)_X$ symmetry are described below.
\begin{itemize}
 \item \underline{\bf{Fermionic decay modes ($Z' \longrightarrow  \overline{\psi}~\psi$)}:}\\
 
 The interactions of $Z'$ with the fermions ($\psi_L, \psi_R$) charged under a given $U(1)_X$ symmetry  can be written as,
\begin{equation}\label{IntZp}
\mathscr{L}_{\text{int}}^{\psi}=-\left( g_{_{\mathtt{L}}}^{\psi} \overline{\psi_{_{\mathtt{L}}}}\gamma^{\mu} \psi_{_{\mathtt{L}}} + g_{_{\mathtt{R}}}^{\psi} \overline{\psi_{_{\mathtt{R}}}}\gamma^{\mu}  \psi_{_{\mathtt{R}}} \right) Z'_{\mu}\,,
\end{equation}
where $g_{l}^{\psi}$ and $g_{r}^{\psi}$ is the coupling of the $Z'$ with left-(right-) handed fermions and given as
\begin{equation}
\begin{split}
g_{_{\mathtt{L}}}^{\psi}&= -(\sin\alpha \cos \theta_{w})gT_{3} + (\sin\alpha \sin\theta_{w})\frac{g' Y_{\psi_{_{\mathtt{L}}}}}{2} + X_{\psi_{_{\mathtt{L}}}}\textsl{g}_{_{x}}\cos\alpha\,,\\
g_{_{\mathtt{R}}}^{\psi}&=  (\sin\alpha \sin\theta_{w})\frac{g' Y_{\psi_{_{\mathtt{R}}}}}{2}+X_{\psi_{_{\mathtt{R}}}} \textsl{g}_{_{x}}\cos\alpha\,.
\end{split}
\end{equation}
For the Dirac fermions, this interaction vertex can be rewritten as
\begin{equation}\label{IntZP2}
\mathscr{L}_{\text{int}}^{\psi}=-\frac{1}{2}\overline{\psi}\gamma^{\mu}(V_{_{\psi}} -\gamma
^{5}A_{_{\psi}})\psi Z'_{\mu}\,,  
\end{equation}
 $\text{ where}\,, \psi = \psi_{_{\mathtt{L}}} + \psi_{_{\mathtt{R}}}\,$. And $V_{_{\psi}}$ and  $A_{_{\psi}}$ are the vector and axial vector couplings defined as, $V_{_{\psi}}= g_{_{\mathtt{L}}}^{\psi}+g_{_{\mathtt{R}}}^{\psi}$, and  $A_{_{\psi}} =g_{_{\mathtt{L}}}^{\psi}-g_{_{\mathtt{R}}}^{\psi}$. The partial decay width of $Z'$, $ \Gamma(Z' \rightarrow \overline{\psi}~\psi )$ into Dirac fermions at tree level can be calculated from Eq. \eqref{IntZP2}, and given as , 
\begin{equation} 
\begin{split}
\Gamma(Z' \rightarrow \overline{\psi}~\psi )&=n_{c}\frac{M_{Z'}}{48\pi}\sqrt{1-\frac{4M^{^{2}}_{\psi}}{{M^{^{2}}_{Z'}}}}  \left[V^{^{2}}_{_{\psi}} +A^{^{2}}_{_{\psi}} +2(V^{^{2}}_{_{\psi}}-2A^{^{2}}_{_{\psi}})\frac{M^{^{2}}_{\psi}}{M^{^{2}}_{Z'}}   \right],
\end{split}
\end{equation}
where $n_{c}$ is the color factor and it is $3$ for quarks and $1$ for leptons. For Majorana fermions, the partial decay width is given as,
\begin{equation}
\Gamma(Z' \rightarrow \overline{\psi_{_{p}}}~\psi_{_{p}} ) = \frac{M_{Z'}}{24\pi} (g_{_{p}}^{\psi})^{2}\left[1-\frac{4M^{^{2}}_{\psi_{_{p}}}}{{M^{^{2}}_{Z'}}} \right]^{3/2}\,,
\end{equation} 
where $g_{_{p}}^{\psi}$ is the coupling to the Majorana fermion with chirality $p$, which could be left~($\mathtt{L}$) or right~($\mathtt{R}$).

\item \underline{\bf{$W^{+}~W^{-}$ decay modes ($Z' \longrightarrow  W^{+}~W^{-}$)~:}}\\

The $Z$ and $Z'$ mix through mass mixing, which results in a vertex involving $Z'$, $W^{+}$ and $W^{-}$.
The partial decay width for $Z' \longrightarrow  W^{+}~W^{-}$ is given by~\cite{Altarelli:1989ff}
\begin{equation}
\begin{split}
\Gamma(Z' \rightarrow W^{+}~W^{-} )&=\frac{(g^{w})^{2}M_{Z'}}{192\pi}\left( \frac{M_{Z'}}{M_{W}}\right)^{4} \left( 1- \frac{4M^{^{2}}_{W}}{M^{^{2}}_{Z'}} \right)^{3/2} \\&~~~~~~~~~~~~~~~~~~\left[ 1+ 20\left(\frac{M_{W}}{M_{Z'}}\right)^{2} + 12\left(\frac{M_{W}}{M_{Z'}}\right)^{4} \right]\,,
\end{split}
\end{equation}
where $g^{w}=g\cos\theta_{w}\sin\alpha$.

\item \underline{\bf{$Z H$ decay modes ($Z' \longrightarrow Z ~H$)~:}}\\

This interaction can be written as 
\begin{equation}
\mathscr{L}_{Z'ZH} = g^{H}Z'_{\mu}Z^{\mu}H \,.
\end{equation}
The partial decay width for $Z' \longrightarrow Z ~H$ is given by
\begin{equation}
\begin{split}
\Gamma(Z' \rightarrow Z~H ) =& \frac{(g^{H})^{2}M_{Z'}}{192\pi M^{^{2}}_{Z}}\left[1-\frac{2M^{^{2}}_{H}-10M^{^{2}}_{Z}}{M^{^{2}}_{Z'}}+\frac{(M^{^{2}}_{H}-M^{^{2}}_{Z})^{2}}{M^{^{4}}_{Z'}} \right]\\
&~~~~~~~~~~~~~~~~~~~~~\left[1-\frac{2(M^{^{2}}_{H}+M^{^{2}}_{Z})}{M^{^{2}}_{Z'}}+\frac{(M^{^{2}}_{H}-M^{^{2}}_{Z})^{2}}{M^{^{4}}_{Z'}} \right]^{1/2}.
\end{split}
\end{equation}
\end{itemize} 
%
%%%%%%%%%%%%%%%%%%%%%%%%%%%%%%%%%%%%%%%%%%%%%%%%%%%%%%%%%%%%%%%%%%%%%%%%%%%%%%%%%%%%%%%%

\section{Feynman diagrams for DM relic density and direct detection}
\label{APn1}

%%%%%%%%%%%%%%%%%%%%%%%%%%%%%%%%%%%%%%%%%%%%%%%%%%%%%%%%%%%%%%%%%%%%%%%%%%%%%%%%%%%%%%%%

\begin{figure*}[ht]
   \centering
   \captionsetup{justification=raggedright}
   \includegraphics[width=0.25\textwidth]{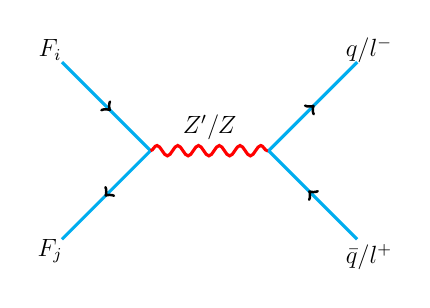}
  \includegraphics[width=0.25\textwidth]{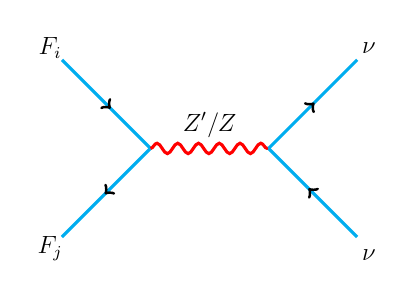}
   \includegraphics[width=0.25\textwidth]{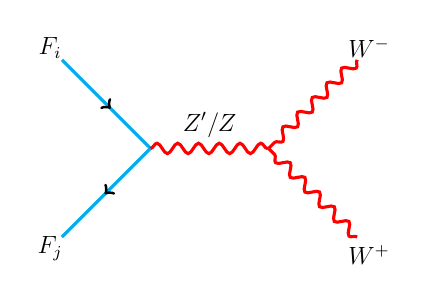}
      \includegraphics[width=0.22\textwidth]{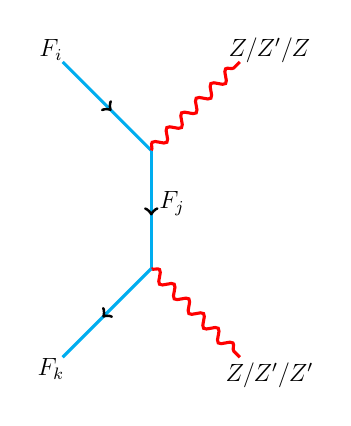}\\
    \includegraphics[width=0.25\textwidth]{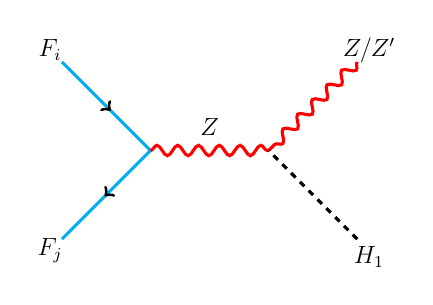}
     \includegraphics[width=0.25\textwidth]{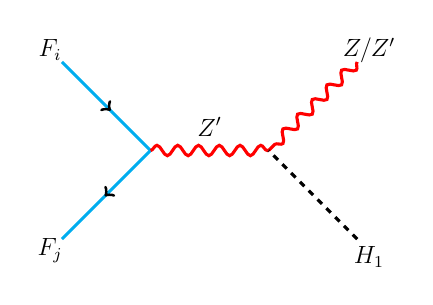}
  \caption{Feynman diagrams that contributes to the relic
density of the fermionic DM, $F_{1}$. Here $q$ represents all quarks and $l$ represents all charged leptons.}
\label{fig:ann}
\end{figure*}

The relic abundance of the DM candidate $F_{1}$, under the benchmark DHC model, is determined by the annihilation and co-annihilation diagrams shown in Fig. \ref{fig:ann} . As previously mentioned, here we have only considered the gauge boson mediated s-channel and the t-channel.

\begin{figure}[ht]
   \centering
   \captionsetup{justification=raggedright}
   \includegraphics[width=0.22\textwidth]{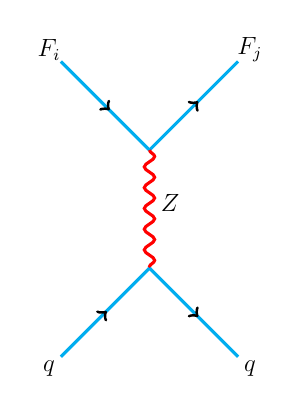}
  \includegraphics[width=0.22\textwidth]{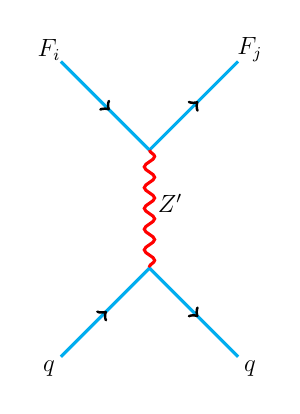}
  \caption{DM-nucleon scattering Feynman diagrams relevant for DM nucleon interactions.}
\label{DDDia}
\end{figure}

 In Fig. \ref{DDDia}, we show the tree-level diagrams that contribute to the spin-dependent DM-nucleon elastic scattering cross-section mediated by the $Z$ boson and $Z'$ boson.

%%%%%%%%%%%%%%%%%%%%%%%%%%%%%%%%%%%%%%%%%%%%%%%%%%%%%%%%%%%%%%%%%%
%%%%%%%%%%%%%%%%%%%%%%%%%%%%%%%%%%%%%%%%%%%%%%%%%%%%%%%%%%%%%%%%%%
\FloatBarrier
\bibliographystyle{utphys}
\bibliography{bibliography}

\providecommand{\href}[2]{#2}\begingroup\raggedright\begin{thebibliography}{10}

\bibitem{Super-Kamiokande:1998kpq}
{\bfseries Super-Kamiokande} Collaboration, Y.~Fukuda {\em et~al.}, ``{Evidence
  for oscillation of atmospheric neutrinos},''
  \href{http://dx.doi.org/10.1103/PhysRevLett.81.1562}{{\em Phys. Rev. Lett.}
  {\bfseries 81} (1998) 1562--1567},
  \href{http://arxiv.org/abs/hep-ex/9807003}{{\ttfamily arXiv:hep-ex/9807003}}.

\bibitem{SNO:2002tuh}
{\bfseries SNO} Collaboration, Q.~R. Ahmad {\em et~al.}, ``{Direct evidence for
  neutrino flavor transformation from neutral current interactions in the
  Sudbury Neutrino Observatory},''
  \href{http://dx.doi.org/10.1103/PhysRevLett.89.011301}{{\em Phys. Rev. Lett.}
  {\bfseries 89} (2002) 011301},
  \href{http://arxiv.org/abs/nucl-ex/0204008}{{\ttfamily
  arXiv:nucl-ex/0204008}}.

\bibitem{Zwicky:1933gu}
F.~Zwicky, ``{Die Rotverschiebung von extragalaktischen Nebeln},''
  \href{http://dx.doi.org/10.1007/s10714-008-0707-4}{{\em Helv. Phys. Acta}
  {\bfseries 6} (1933) 110--127}.

\bibitem{Rubin:1970zza}
V.~C. Rubin and W.~K. Ford, Jr., ``{Rotation of the Andromeda Nebula from a
  Spectroscopic Survey of Emission Regions},''
  \href{http://dx.doi.org/10.1086/150317}{{\em Astrophys. J.} {\bfseries 159}
  (1970) 379--403}.

\bibitem{Rubin:1980zd}
V.~C. Rubin, N.~Thonnard, and W.~K. Ford, Jr., ``{Rotational properties of 21
  SC galaxies with a large range of luminosities and radii, from NGC 4605 /R =
  4kpc/ to UGC 2885 /R = 122 kpc/},''
  \href{http://dx.doi.org/10.1086/158003}{{\em Astrophys. J.} {\bfseries 238}
  (1980) 471}.

\bibitem{Planck:2018vyg}
{\bfseries Planck} Collaboration, N.~Aghanim {\em et~al.}, ``{Planck 2018
  results. VI. Cosmological parameters},''
  \href{http://dx.doi.org/10.1051/0004-6361/201833910}{{\em Astron. Astrophys.}
  {\bfseries 641} (2020) A6}, \href{http://arxiv.org/abs/1807.06209}{{\ttfamily
  arXiv:1807.06209 [astro-ph.CO]}}. [Erratum: Astron.Astrophys. 652, C4
  (2021)].

\bibitem{He:1990pn}
X.~G. He, G.~C. Joshi, H.~Lew, and R.~R. Volkas, ``{NEW Z-prime
  PHENOMENOLOGY},'' \href{http://dx.doi.org/10.1103/PhysRevD.43.R22}{{\em Phys.
  Rev. D} {\bfseries 43} (1991) 22--24}.

\bibitem{Ma:1997nq}
E.~Ma, ``{Gauged B - 3L(tau) and radiative neutrino masses},''
  \href{http://dx.doi.org/10.1016/S0370-2693(98)00599-1}{{\em Phys. Lett. B}
  {\bfseries 433} (1998) 74--81},
  \href{http://arxiv.org/abs/hep-ph/9709474}{{\ttfamily arXiv:hep-ph/9709474}}.

\bibitem{Appelquist:2002mw}
T.~Appelquist, B.~A. Dobrescu, and A.~R. Hopper, ``{Nonexotic Neutral Gauge
  Bosons},'' \href{http://dx.doi.org/10.1103/PhysRevD.68.035012}{{\em Phys.
  Rev. D} {\bfseries 68} (2003) 035012},
  \href{http://arxiv.org/abs/hep-ph/0212073}{{\ttfamily arXiv:hep-ph/0212073}}.

\bibitem{Montero:2007cd}
J.~C. Montero and V.~Pleitez, ``{Gauging U(1) symmetries and the number of
  right-handed neutrinos},''
  \href{http://dx.doi.org/10.1016/j.physletb.2009.03.065}{{\em Phys. Lett. B}
  {\bfseries 675} (2009) 64--68},
  \href{http://arxiv.org/abs/0706.0473}{{\ttfamily arXiv:0706.0473 [hep-ph]}}.

\bibitem{Lee:2010hf}
H.-S. Lee and E.~Ma, ``{Gauged $B-x_i L$ origin of $R$ Parity and its
  implications},'' \href{http://dx.doi.org/10.1016/j.physletb.2010.04.032}{{\em
  Phys. Lett. B} {\bfseries 688} (2010) 319--322},
  \href{http://arxiv.org/abs/1001.0768}{{\ttfamily arXiv:1001.0768 [hep-ph]}}.

\bibitem{Ma:2014qra}
E.~Ma and R.~Srivastava, ``{Dirac or inverse seesaw neutrino masses with $B-L$
  gauge symmetry and $S_3$ flavor symmetry},''
  \href{http://dx.doi.org/10.1016/j.physletb.2014.12.049}{{\em Phys. Lett. B}
  {\bfseries 741} (2015) 217--222},
  \href{http://arxiv.org/abs/1411.5042}{{\ttfamily arXiv:1411.5042 [hep-ph]}}.

\bibitem{Ma:2015raa}
E.~Ma and R.~Srivastava, ``{Dirac or inverse seesaw neutrino masses from gauged
  $B–L$ symmetry},'' \href{http://dx.doi.org/10.1142/S0217732315300207}{{\em
  Mod. Phys. Lett. A} {\bfseries 30} no.~26, (2015) 1530020},
  \href{http://arxiv.org/abs/1504.00111}{{\ttfamily arXiv:1504.00111
  [hep-ph]}}.

\bibitem{Ma:2015mjd}
E.~Ma, N.~Pollard, R.~Srivastava, and M.~Zakeri, ``{Gauge $B-L$ Model with
  Residual $Z_3$ Symmetry},''
  \href{http://dx.doi.org/10.1016/j.physletb.2015.09.010}{{\em Phys. Lett. B}
  {\bfseries 750} (2015) 135--138},
  \href{http://arxiv.org/abs/1507.03943}{{\ttfamily arXiv:1507.03943
  [hep-ph]}}.

\bibitem{Das:2016zue}
A.~Das, S.~Oda, N.~Okada, and D.-s. Takahashi, ``{Classically conformal U(1)'
  extended standard model, electroweak vacuum stability, and LHC Run-2
  bounds},'' \href{http://dx.doi.org/10.1103/PhysRevD.93.115038}{{\em Phys.
  Rev. D} {\bfseries 93} no.~11, (2016) 115038},
  \href{http://arxiv.org/abs/1605.01157}{{\ttfamily arXiv:1605.01157
  [hep-ph]}}.

\bibitem{Bonilla:2017lsq}
C.~Bonilla, T.~Modak, R.~Srivastava, and J.~W.~F. Valle, ``{$U(1)_{B_3-3L_\mu}$
  gauge symmetry as a simple description of $b\to s$ anomalies},''
  \href{http://dx.doi.org/10.1103/PhysRevD.98.095002}{{\em Phys. Rev. D}
  {\bfseries 98} no.~9, (2018) 095002},
  \href{http://arxiv.org/abs/1705.00915}{{\ttfamily arXiv:1705.00915
  [hep-ph]}}.

\bibitem{Alonso:2017uky}
R.~Alonso, P.~Cox, C.~Han, and T.~T. Yanagida, ``{Flavoured $B-L$ local
  symmetry and anomalous rare $B$ decays},''
  \href{http://dx.doi.org/10.1016/j.physletb.2017.10.027}{{\em Phys. Lett. B}
  {\bfseries 774} (2017) 643--648},
  \href{http://arxiv.org/abs/1705.03858}{{\ttfamily arXiv:1705.03858
  [hep-ph]}}.

\bibitem{Jana:2019mez}
S.~Jana, P.~K. Vishnu, and S.~Saad, ``{Minimal dirac neutrino mass models from
  $\hbox {U}(1)_{\mathrm{R}}$ gauge symmetry and left\textendash{}right
  asymmetry at colliders},''
  \href{http://dx.doi.org/10.1140/epjc/s10052-019-7441-9}{{\em Eur. Phys. J. C}
  {\bfseries 79} no.~11, (2019) 916},
  \href{http://arxiv.org/abs/1904.07407}{{\ttfamily arXiv:1904.07407
  [hep-ph]}}.

\bibitem{DeRomeri:2023ytt}
V.~De~Romeri, A.~Majumdar, D.~K. Papoulias, and R.~Srivastava, ``{XENONnT and
  LUX-ZEPLIN constraints on DSNB-boosted dark matter},''
  \href{http://dx.doi.org/10.1088/1475-7516/2024/03/028}{{\em JCAP} {\bfseries
  03} (2024) 028}, \href{http://arxiv.org/abs/2309.04117}{{\ttfamily
  arXiv:2309.04117 [hep-ph]}}.

\bibitem{Mandal:2023oyh}
S.~Mandal, H.~Prajapati, and R.~Srivastava, ``{$B-L$ model in light of the CDF
  II result},'' \href{http://arxiv.org/abs/2301.01522}{{\ttfamily
  arXiv:2301.01522 [hep-ph]}}.

\bibitem{Ghosh:2024cxi}
D.~K. Ghosh, P.~Ghosh, S.~Jeesun, and R.~Srivastava, ``{Neff at CMB challenges
  U(1)X light gauge boson scenarios},''
  \href{http://dx.doi.org/10.1103/PhysRevD.110.075032}{{\em Phys. Rev. D}
  {\bfseries 110} no.~7, (2024) 075032},
  \href{http://arxiv.org/abs/2404.10077}{{\ttfamily arXiv:2404.10077
  [hep-ph]}}.

\bibitem{Adler:1969gk}
S.~L. Adler, ``{Axial vector vertex in spinor electrodynamics},''
  \href{http://dx.doi.org/10.1103/PhysRev.177.2426}{{\em Phys. Rev.} {\bfseries
  177} (1969) 2426--2438}.

\bibitem{Bardeen:1969md}
W.~A. Bardeen, ``{Anomalous Ward identities in spinor field theories},''
  \href{http://dx.doi.org/10.1103/PhysRev.184.1848}{{\em Phys. Rev.} {\bfseries
  184} (1969) 1848--1857}.

\bibitem{Bell:1969ts}
J.~S. Bell and R.~Jackiw, ``{A PCAC puzzle: $\pi^0 \to \gamma \gamma$ in the
  $\sigma$ model},'' \href{http://dx.doi.org/10.1007/BF02823296}{{\em Nuovo
  Cim. A} {\bfseries 60} (1969) 47--61}.

\bibitem{Delbourgo:1972xb}
R.~Delbourgo and A.~Salam, ``{The gravitational correction to pcac},''
  \href{http://dx.doi.org/10.1016/0370-2693(72)90825-8}{{\em Phys. Lett. B}
  {\bfseries 40} (1972) 381--382}.

\bibitem{Alvarez-Gaume:1983ihn}
L.~Alvarez-Gaume and E.~Witten, ``{Gravitational Anomalies},''
  \href{http://dx.doi.org/10.1016/0550-3213(84)90066-X}{{\em Nucl. Phys. B}
  {\bfseries 234} (1984) 269}.

\bibitem{Witten:1982fp}
E.~Witten, ``{An SU(2) Anomaly},''
  \href{http://dx.doi.org/10.1016/0370-2693(82)90728-6}{{\em Phys. Lett. B}
  {\bfseries 117} (1982) 324--328}.

\bibitem{Geng:1989tcu}
C.~Q. Geng and R.~E. Marshak, ``{Uniqueness of Quark and Lepton Representations
  in the Standard Model From the Anomalies Viewpoint},''
  \href{http://dx.doi.org/10.1103/PhysRevD.39.693}{{\em Phys. Rev. D}
  {\bfseries 39} (1989) 693}.

\bibitem{Minahan:1989vd}
J.~A. Minahan, P.~Ramond, and R.~C. Warner, ``{A Comment on Anomaly
  Cancellation in the Standard Model},''
  \href{http://dx.doi.org/10.1103/PhysRevD.41.715}{{\em Phys. Rev. D}
  {\bfseries 41} (1990) 715}.

\bibitem{Allanach:2018vjg}
B.~C. Allanach, J.~Davighi, and S.~Melville, ``{An Anomaly-free Atlas: charting
  the space of flavour-dependent gauged $U(1)$ extensions of the Standard
  Model},'' \href{http://dx.doi.org/10.1007/JHEP02(2019)082}{{\em JHEP}
  {\bfseries 02} (2019) 082}, \href{http://arxiv.org/abs/1812.04602}{{\ttfamily
  arXiv:1812.04602 [hep-ph]}}. [Erratum: JHEP 08, 064 (2019)].

\bibitem{Allanach:2022blr}
B.~Allanach and E.~Loisa, ``{Flavonstrahlung in the B$_{3}$\ensuremath{-}
  L$_{2}$Z' model at current and future colliders},''
  \href{http://dx.doi.org/10.1007/JHEP03(2023)253}{{\em JHEP} {\bfseries 03}
  (2023) 253}, \href{http://arxiv.org/abs/2212.07440}{{\ttfamily
  arXiv:2212.07440 [hep-ph]}}.

\bibitem{Farzan:2015doa}
Y.~Farzan, ``{A model for large non-standard interactions of neutrinos leading
  to the LMA-Dark solution},''
  \href{http://dx.doi.org/10.1016/j.physletb.2015.07.015}{{\em Phys. Lett. B}
  {\bfseries 748} (2015) 311--315},
  \href{http://arxiv.org/abs/1505.06906}{{\ttfamily arXiv:1505.06906
  [hep-ph]}}.

\bibitem{DeRomeri:2024dbv}
V.~De~Romeri, D.~K. Papoulias, and C.~A. Ternes, ``{Light vector mediators at
  direct detection experiments},''
  \href{http://dx.doi.org/10.1007/JHEP05(2024)165}{{\em JHEP} {\bfseries 05}
  (2024) 165}, \href{http://arxiv.org/abs/2402.05506}{{\ttfamily
  arXiv:2402.05506 [hep-ph]}}.

\bibitem{AtzoriCorona:2022moj}
M.~Atzori~Corona, M.~Cadeddu, N.~Cargioli, F.~Dordei, C.~Giunti, Y.~F. Li,
  E.~Picciau, C.~A. Ternes, and Y.~Y. Zhang, ``{Probing light mediators and (g
  \ensuremath{-} 2)$_{μ}$ through detection of coherent elastic neutrino
  nucleus scattering at COHERENT},''
  \href{http://dx.doi.org/10.1007/JHEP05(2022)109}{{\em JHEP} {\bfseries 05}
  (2022) 109}, \href{http://arxiv.org/abs/2202.11002}{{\ttfamily
  arXiv:2202.11002 [hep-ph]}}.

\bibitem{Rahul:2024N}
D.~Malayaja, O.~Popov, and R.~Srivastava, ``{Model for gauged Baryon and Lepton
  number symmetries},''.

\bibitem{Bento:2023flt}
M.~P. Bento, H.~E. Haber, and J.~a.~P. Silva, ``{Classes of complete dark
  photon models constrained by Z-physics},''
  \href{http://dx.doi.org/10.1016/j.physletb.2024.138501}{{\em P,hys. Lett. B}
  {\bfseries 850} (2024) 138501},
  \href{http://arxiv.org/abs/2311.04976}{{\ttfamily arXiv:2311.04976
  [hep-ph]}}.

\bibitem{Bento:2023weq}
M.~P. Bento, H.~E. Haber, and J.~a.~P. Silva, ``{Tree-level Unitarity in $
  \textrm{SU}{(2)}_L\times \textrm{U}{(1)}_Y\times \textrm{U}{(1)}_{Y^{\prime
  }} $ Models},'' \href{http://dx.doi.org/10.1007/JHEP10(2023)083}{{\em JHEP}
  {\bfseries 10} (2023) 083}, \href{http://arxiv.org/abs/2306.01836}{{\ttfamily
  arXiv:2306.01836 [hep-ph]}}.

\bibitem{ParticleDataGroup:2020ssz}
{\bfseries Particle Data Group} Collaboration, P.~A. Zyla {\em et~al.},
  ``{Review of Particle Physics},''
  \href{http://dx.doi.org/10.1093/ptep/ptaa104}{{\em PTEP} {\bfseries 2020}
  no.~8, (2020) 083C01}.

\bibitem{CentellesChulia:2018gwr}
S.~Centelles~Chuli\'a, R.~Srivastava, and J.~W.~F. Valle, ``{Seesaw roadmap to
  neutrino mass and dark matter},''
  \href{http://dx.doi.org/10.1016/j.physletb.2018.03.046}{{\em Phys. Lett. B}
  {\bfseries 781} (2018) 122--128},
  \href{http://arxiv.org/abs/1802.05722}{{\ttfamily arXiv:1802.05722
  [hep-ph]}}.

\bibitem{ATLAS:2019erb}
{\bfseries ATLAS} Collaboration, G.~Aad {\em et~al.}, ``{Search for high-mass
  dilepton resonances using 139 fb$^{-1}$ of $pp$ collision data collected at
  $\sqrt{s}=$13 TeV with the ATLAS detector},''
  \href{http://dx.doi.org/10.1016/j.physletb.2019.07.016}{{\em Phys. Lett. B}
  {\bfseries 796} (2019) 68--87},
  \href{http://arxiv.org/abs/1903.06248}{{\ttfamily arXiv:1903.06248
  [hep-ex]}}.

\bibitem{CMS:2021ctt}
{\bfseries CMS} Collaboration, A.~M. Sirunyan {\em et~al.}, ``{Search for
  resonant and nonresonant new phenomena in high-mass dilepton final states at
  $ \sqrt{s} $ = 13 TeV},''
  \href{http://dx.doi.org/10.1007/JHEP07(2021)208}{{\em JHEP} {\bfseries 07}
  (2021) 208}, \href{http://arxiv.org/abs/2103.02708}{{\ttfamily
  arXiv:2103.02708 [hep-ex]}}.

\bibitem{Mahanthappa:1991pw}
K.~T. Mahanthappa and P.~K. Mohapatra, ``{Limits on mixing angle and mass of
  Z-prime using Delta rho and atomic parity violation},''
  \href{http://dx.doi.org/10.1103/PhysRevD.43.3093}{{\em Phys. Rev. D}
  {\bfseries 43} (1991) 3093}. [Erratum: Phys.Rev.D 44, 1616 (1991)].

\bibitem{Diener:2011jt}
R.~Diener, S.~Godfrey, and I.~Turan, ``{Constraining Extra Neutral Gauge Bosons
  with Atomic Parity Violation Measurements},''
  \href{http://dx.doi.org/10.1103/PhysRevD.86.115017}{{\em Phys. Rev. D}
  {\bfseries 86} (2012) 115017},
  \href{http://arxiv.org/abs/1111.4566}{{\ttfamily arXiv:1111.4566 [hep-ph]}}.

\bibitem{Accomando:2015cfa}
E.~Accomando, A.~Belyaev, J.~Fiaschi, K.~Mimasu, S.~Moretti, and
  C.~Shepherd-Themistocleous, ``{Forward-backward asymmetry as a discovery tool
  for Z' bosons at the LHC},''
  \href{http://dx.doi.org/10.1007/JHEP01(2016)127}{{\em JHEP} {\bfseries 01}
  (2016) 127}, \href{http://arxiv.org/abs/1503.02672}{{\ttfamily
  arXiv:1503.02672 [hep-ph]}}.

\bibitem{LZCollaboration:2024lux}
{\bfseries LZ Collaboration} Collaboration, J.~Aalbers {\em et~al.}, ``{Dark
  Matter Search Results from 4.2 Tonne-Years of Exposure of the LUX-ZEPLIN (LZ)
  Experiment},'' \href{http://arxiv.org/abs/2410.17036}{{\ttfamily
  arXiv:2410.17036 [hep-ex]}}.

\bibitem{PandaX:2024qfu}
{\bfseries PandaX} Collaboration, Z.~Bo {\em et~al.}, ``{Dark Matter Search
  Results from 1.54 Tonne$\cdot$Year Exposure of PandaX-4T},''
  \href{http://arxiv.org/abs/2408.00664}{{\ttfamily arXiv:2408.00664
  [hep-ex]}}.

\bibitem{XENONCollaboration:2023orw}
{\bfseries (XENON Collaboration)\textdagger{}\textdagger{}, XENON}
  Collaboration, E.~Aprile {\em et~al.}, ``{First Dark Matter Search with
  Nuclear Recoils from the XENONnT Experiment},''
  \href{http://dx.doi.org/10.1103/PhysRevLett.131.041003}{{\em Phys. Rev.
  Lett.} {\bfseries 131} no.~4, (2023) 041003},
  \href{http://arxiv.org/abs/2303.14729}{{\ttfamily arXiv:2303.14729
  [hep-ex]}}.

\bibitem{IceCube:2016dgk}
{\bfseries IceCube} Collaboration, M.~G. Aartsen {\em et~al.}, ``{Search for
  annihilating dark matter in the Sun with 3 years of IceCube data},''
  \href{http://dx.doi.org/10.1140/epjc/s10052-017-4689-9}{{\em Eur. Phys. J. C}
  {\bfseries 77} no.~3, (2017) 146},
  \href{http://arxiv.org/abs/1612.05949}{{\ttfamily arXiv:1612.05949
  [astro-ph.HE]}}. [Erratum: Eur.Phys.J.C 79, 214 (2019)].

\bibitem{ANTARES:2016xuh}
{\bfseries ANTARES} Collaboration, S.~Adrian-Martinez {\em et~al.}, ``{Limits
  on Dark Matter Annihilation in the Sun using the ANTARES Neutrino
  Telescope},'' \href{http://dx.doi.org/10.1016/j.physletb.2016.05.019}{{\em
  Phys. Lett. B} {\bfseries 759} (2016) 69--74},
  \href{http://arxiv.org/abs/1603.02228}{{\ttfamily arXiv:1603.02228
  [astro-ph.HE]}}.

\bibitem{Frankiewicz:2015zma}
{\bfseries Super-Kamiokande} Collaboration, K.~Frankiewicz, ``{Searching for
  Dark Matter Annihilation into Neutrinos with Super-Kamiokande},'' in {\em
  {Meeting of the APS Division of Particles and Fields}}.
\newblock 10, 2015.
\newblock \href{http://arxiv.org/abs/1510.07999}{{\ttfamily arXiv:1510.07999
  [hep-ex]}}.

\bibitem{Majumdar:2024dms}
A.~Majumdar, D.~K. Papoulias, H.~Prajapati, and R.~Srivastava, ``{Constraining
  low scale Dark Hypercharge symmetry at spallation, reactor and Dark Matter
  direct detection experiments},''
  \href{http://arxiv.org/abs/2411.04197}{{\ttfamily arXiv:2411.04197
  [hep-ph]}}.

\bibitem{Krasznahorkay:2015iga}
A.~J. Krasznahorkay {\em et~al.}, ``{Observation of Anomalous Internal Pair
  Creation in Be8 : A Possible Indication of a Light, Neutral Boson},''
  \href{http://dx.doi.org/10.1103/PhysRevLett.116.042501}{{\em Phys. Rev.
  Lett.} {\bfseries 116} no.~4, (2016) 042501},
  \href{http://arxiv.org/abs/1504.01527}{{\ttfamily arXiv:1504.01527
  [nucl-ex]}}.

\bibitem{Krasznahorkay:2021joi}
A.~J. Krasznahorkay, M.~Csatl\'os, L.~Csige, J.~Guly\'as, A.~Krasznahorkay,
  B.~M. Nyak\'o, I.~Rajta, J.~Tim\'ar, I.~Vajda, and N.~J. Sas, ``{New anomaly
  observed in He4 supports the existence of the hypothetical X17 particle},''
  \href{http://dx.doi.org/10.1103/PhysRevC.104.044003}{{\em Phys. Rev. C}
  {\bfseries 104} no.~4, (2021) 044003},
  \href{http://arxiv.org/abs/2104.10075}{{\ttfamily arXiv:2104.10075
  [nucl-ex]}}.

\bibitem{Krasznahorkay:2022pxs}
A.~J. Krasznahorkay {\em et~al.}, ``{New anomaly observed in C12 supports the
  existence and the vector character of the hypothetical X17 boson},''
  \href{http://dx.doi.org/10.1103/PhysRevC.106.L061601}{{\em Phys. Rev. C}
  {\bfseries 106} no.~6, (2022) L061601},
  \href{http://arxiv.org/abs/2209.10795}{{\ttfamily arXiv:2209.10795
  [nucl-ex]}}.

\bibitem{Barducci:2022lqd}
D.~Barducci and C.~Toni, ``{An updated view on the ATOMKI nuclear anomalies},''
  \href{http://dx.doi.org/10.1007/JHEP02(2023)154}{{\em JHEP} {\bfseries 02}
  (2023) 154}, \href{http://arxiv.org/abs/2212.06453}{{\ttfamily
  arXiv:2212.06453 [hep-ph]}}. [Erratum: JHEP 07, 168 (2023)].

\bibitem{Bossi:2025ptv}
F.~Bossi {\em et~al.}, ``{Search for a new 17 MeV resonance via $e^+e^-$
  annihilation with the PADME Experiment},''
  \href{http://arxiv.org/abs/2505.24797}{{\ttfamily arXiv:2505.24797
  [hep-ex]}}.

\bibitem{Arias-Aragon:2025wdt}
F.~Arias-Arag\'on, G.~G. di~Cortona, E.~Nardi, and C.~Toni, ``{Combined
  Evidence for the $X_{17}$ Boson After PADME Results on Resonant Production in
  Positron Annihilation},'' \href{http://arxiv.org/abs/2504.11439}{{\ttfamily
  arXiv:2504.11439 [hep-ph]}}.

\bibitem{KTeV:2006pwx}
{\bfseries KTeV} Collaboration, E.~Abouzaid {\em et~al.}, ``{Measurement of the
  Rare Decay $\pi^0 \to e^+ e^-$},''
  \href{http://dx.doi.org/10.1103/PhysRevD.75.012004}{{\em Phys. Rev. D}
  {\bfseries 75} (2007) 012004},
  \href{http://arxiv.org/abs/hep-ex/0610072}{{\ttfamily arXiv:hep-ex/0610072}}.

\bibitem{Parker:2018vye}
R.~H. Parker, C.~Yu, W.~Zhong, B.~Estey, and H.~M{\"u}ller, ``{Measurement of
  the fine-structure constant as a test of the Standard Model},''
  \href{http://dx.doi.org/10.1126/science.aap7706}{{\em Science} {\bfseries
  360} (2018) 191}, \href{http://arxiv.org/abs/1812.04130}{{\ttfamily
  arXiv:1812.04130 [physics.atom-ph]}}.

\bibitem{Belle-II:2023esi}
{\bfseries Belle-II} Collaboration, I.~Adachi {\em et~al.}, ``{Evidence for
  B+\textrightarrow{}K+\ensuremath{\nu}\ensuremath{\nu}\textasciimacron{}
  decays},'' \href{http://dx.doi.org/10.1103/PhysRevD.109.112006}{{\em Phys.
  Rev. D} {\bfseries 109} no.~11, (2024) 112006},
  \href{http://arxiv.org/abs/2311.14647}{{\ttfamily arXiv:2311.14647
  [hep-ex]}}.

\bibitem{Staub:2015kfa}
F.~Staub, ``{Exploring new models in all detail with SARAH},''
  \href{http://dx.doi.org/10.1155/2015/840780}{{\em Adv. High Energy Phys.}
  {\bfseries 2015} (2015) 840780},
  \href{http://arxiv.org/abs/1503.04200}{{\ttfamily arXiv:1503.04200
  [hep-ph]}}.

\bibitem{Goodsell:2017pdq}
M.~D. Goodsell, S.~Liebler, and F.~Staub, ``{Generic calculation of two-body
  partial decay widths at the full one-loop level},''
  \href{http://dx.doi.org/10.1140/epjc/s10052-017-5259-x}{{\em Eur. Phys. J. C}
  {\bfseries 77} no.~11, (2017) 758},
  \href{http://arxiv.org/abs/1703.09237}{{\ttfamily arXiv:1703.09237
  [hep-ph]}}.

\bibitem{Porod:2003um}
W.~Porod, ``{SPheno, a program for calculating supersymmetric spectra, SUSY
  particle decays and SUSY particle production at e+ e- colliders},''
  \href{http://dx.doi.org/10.1016/S0010-4655(03)00222-4}{{\em Comput. Phys.
  Commun.} {\bfseries 153} (2003) 275--315},
  \href{http://arxiv.org/abs/hep-ph/0301101}{{\ttfamily arXiv:hep-ph/0301101}}.

\bibitem{Belanger:2020gnr}
G.~Belanger, A.~Mjallal, and A.~Pukhov, ``{Recasting direct detection limits
  within micrOMEGAs and implication for non-standard Dark Matter scenarios},''
  \href{http://dx.doi.org/10.1140/epjc/s10052-021-09012-z}{{\em Eur. Phys. J.
  C} {\bfseries 81} no.~3, (2021) 239},
  \href{http://arxiv.org/abs/2003.08621}{{\ttfamily arXiv:2003.08621
  [hep-ph]}}.

\bibitem{Belanger:2014vza}
G.~B\'elanger, F.~Boudjema, A.~Pukhov, and A.~Semenov, ``{micrOMEGAs4.1: two
  dark matter candidates},''
  \href{http://dx.doi.org/10.1016/j.cpc.2015.03.003}{{\em Comput. Phys.
  Commun.} {\bfseries 192} (2015) 322--329},
  \href{http://arxiv.org/abs/1407.6129}{{\ttfamily arXiv:1407.6129 [hep-ph]}}.

\bibitem{Alwall:2014hca}
J.~Alwall, R.~Frederix, S.~Frixione, V.~Hirschi, F.~Maltoni, O.~Mattelaer,
  H.~S. Shao, T.~Stelzer, P.~Torrielli, and M.~Zaro, ``{The automated
  computation of tree-level and next-to-leading order differential cross
  sections, and their matching to parton shower simulations},''
  \href{http://dx.doi.org/10.1007/JHEP07(2014)079}{{\em JHEP} {\bfseries 07}
  (2014) 079}, \href{http://arxiv.org/abs/1405.0301}{{\ttfamily arXiv:1405.0301
  [hep-ph]}}.

\bibitem{Altarelli:1989ff}
G.~Altarelli, B.~Mele, and M.~Ruiz-Altaba, ``{Searching for New Heavy Vector
  Bosons in $p \bar{p}$ Colliders},''
  \href{http://dx.doi.org/10.1007/BF01556677}{{\em Z. Phys. C} {\bfseries 45}
  (1989) 109}. [Erratum: Z.Phys.C 47, 676 (1990)].

\end{thebibliography}\endgroup

\end{document}